\documentclass[sigconf]{acmart}

\usepackage{multirow}

\usepackage{amssymb}
\usepackage{amsmath,amsthm,mathtools} 
\usepackage{algorithm} 
\usepackage{algpseudocode}
\usepackage{graphicx}
\usepackage{varwidth} 
\usepackage{float}        
\usepackage{microtype}    
\usepackage{subcaption}
\usepackage{enumitem}
\usepackage{makecell}

\AtBeginDocument{%
  }

\copyrightyear{2026}
\acmYear{2026}
\setcopyright{cc}
\setcctype{by}
\acmConference[RecSys '26]{20th ACM Conference on Recommender Systems}{September 27-October 02, 2026}{Minneapolis, MN, USA}
\acmBooktitle{20th ACM Conference on Recommender Systems (RecSys '26), September 27-October 02, 2026, Minneapolis, MN, USA}
\acmDOI{10.1145/3773078.3831756}
\acmISBN{979-8-4007-2284-4/2026/09}

\begin{document}

\title{Not All Branches Are Equal: Adaptive Semantic ID Construction for Generative Recommendation}

\author{Guy Hadad}
\orcid{0009-0004-6264-5127}
\email{guyhada@post.bgu.ac.il}
\affiliation{%
  \institution{Ben-Gurion University of the Negev}
  \city{Beer Sheva}
  \country{Israel}
}

\author{Haggai Roitman}
\email{haggair@gmail.com}
\affiliation{%
  \institution{Ben-Gurion University of the Negev}
  \city{Beer Sheva}
  \country{Israel}
}

\author{Bracha Shapira}
\orcid{0000-0003-4943-9324}
\email{bshapira@bgu.ac.il}
\affiliation{%
  \institution{Ben-Gurion University of the Negev}
  \city{Beer Sheva}
  \country{Israel}
}
\renewcommand{\shortauthors}{Hadad et al.}

\begin{abstract}

Generative recommendation models represent items as sequences of discrete tokens known as semantic IDs. Existing approaches derive these identifiers using residual or hierarchical vector quantization, assuming that tokens carry independent semantic meaning across quantization levels. 
Revisiting this assumption, we argue that semantic IDs should behave as hierarchical paths where each token's meaning is conditioned on its prefix. 
This highlights a mismatch between conventional fixed-codebook quantization and the hierarchical structure implicitly learned by generative recommenders. Motivated by this insight, we propose \textit{Adaptive Silhouette Tree} (AST), a top-down divisive clustering method that constructs semantic IDs as paths within a tree structure. Unlike prior methods with fixed branching factors, AST dynamically determines the number of branches at each node by maximizing the silhouette coefficient, enabling finer partitions in dense regions and compact groupings in homogeneous ones. We further introduce constrained variants that enforce minimum probability mass during splitting to improve supervision for long-tail items. Extensive experiments on five Amazon Review datasets of varying sizes and domains show that AST 
significantly outperforms strong baselines. 
These results demonstrate that aligning semantic ID construction with the inherent hierarchical structure of generative recommenders leads to more effective recommendations.

\end{abstract}

\ccsdesc[500]{Information systems~Recommender systems}

\keywords{Recommendation, Semantic IDs, Tokenization}


\maketitle

\section{Introduction}

Generative recommendation has recently emerged as a powerful paradigm for modern recommender systems, reframing recommendation as a sequence generation problem rather than a direct scoring or ranking task \citep{rajput2023recommender, wang2024learnable, zheng2024adapting}. In this formulation, items are represented as sequences of discrete tokens, commonly referred to as semantic IDs (SIDs), and recommendation is performed by generating the next token sequence conditioned on user history. This approach enables seamless integration with autoregressive language models, providing scalable decoding, compact representations, and a natural way to leverage rich item content such as textual descriptions \cite{wu2024survey, zhao2024recommender,hadad2026encoderec}.

Generative recommendation stems from two research lines. The first, generative modeling in high-dimensional modalities like images and audio, uses vector quantization and neural tokenization \cite{lee2022autoregressive, tian2024visual, borsos2023audiolm, rubenstein2023audiopalm} to represent signals as discrete autoregressive sequences. The second, generative retrieval \cite{tay2022transformer, zeng2024scalable}, models document identifiers as token sequences generated by sequence models, unifying retrieval and generation. These methods are vital for large-scale systems. Platforms like YouTube or Spotify manage hundreds of millions of items, requiring efficient indexing. Traditional pipelines use complex multi-stage retrieval and ranking. Generative recommendation offers an alternative by directly generating item identifiers, simplifying system design and enabling richer modeling of user preferences and content \cite{rajput2023recommender,he2025plum,denadai2026unifiedlanguagemodellarge}. However, its effectiveness depends on how items are tokenized into identifiers. 


While central to the field, the construction of semantic IDs in existing methods often relies on techniques adapted from other domains, which may not be fully tailored to the specific requirements of recommendation. 
Most approaches are based on vector quantization or residual-based clustering of item embeddings~\cite{kong2025minionerec, he2025plum, rajput2023recommender, tang2025reaseq, wang2024learnable,ju2025generative,zhu2024cost,pang2025higr,vallaeys2025qinco2}, implicitly assuming each token carries independent semantic meaning, or treating quantization as the primary objective with codebooks used for indexing. In this work, we revisit this assumption and propose a different perspective: \textit{semantic IDs should be viewed as paths in a tree, where the meaning of each token is fully conditioned on its prefix}. Through empirical analysis, we show that tokens do not exhibit independent semantics, but instead derive their meaning from their position within a hierarchy. 

Motivated by this observation, we propose the \textit{Adaptive Silhouette Tree} (AST), a method for constructing semantic IDs using top-down hierarchical clustering with dynamic branching. At each node, AST adaptively selects the branching factor based on the local structure of the data, enabling finer partitions in dense and diverse regions while maintaining coarser partitions in more homogeneous areas. To address 
insufficient supervision for rare prefixes, we introduce constraints on the minimum probability mass assigned to each branch. We enforce these constraints using mixed-integer linear programming (MILP), guaranteeing sufficient supervision for all prefixes and eliminating degenerate low-support paths.


We provide a comprehensive evaluation of our approach, analyzing its theoretical properties and conducting extensive experiments across multiple Amazon Review datasets. Our study includes comparisons with existing semantic ID construction methods, as well as analyses of head versus tail performance, sensitivity to key hyperparameters, and ablations isolating each component’s contribution. Across all datasets, our method achieves an average improvement of at least 13\% over the strongest baseline across all metrics, highlighting the advantage of modeling semantic IDs as hierarchical paths rather than independent tokens. We conclude with a discussion of limitations and future directions.

\section{Related Work}

The success of autoregressive models in text generation \cite{vaswani2017attention, raffel2020exploring, yang2025qwen3} has inspired similar approaches for continuous data. In images and audio, this is done by transforming signals into discrete (token) sequences via quantization methods like VQ-VAE \cite{van2017neural}, RQ-VAE \cite{lee2022autoregressive}, and variants \cite{razavi2019generating, esser2021taming, yu2023language, borsos2023audiolm, defossez2022high}, highlighting the importance of tokenizer design.
In parallel, generative retrieval models such as GENRE \cite{de2020autoregressive}, DSI \cite{tay2022transformer}, and NCI \cite{wang2022neural} treat retrieval as sequence generation over tokenized identifiers. Building on this, recommendation has shifted toward \emph{generative recommendation}, where items are encoded as semantic ID sequences and predicted autoregressively. This replaces large-scale classification with sequence generation, making tokenization a central design challenge. 

A dominant line of work in generative recommendation constructs semantic IDs through residual quantization, where the quantization objective itself drives the identifier construction and the resulting index is typically formed from the selected codebook entries at each level. TIGER \cite{rajput2023recommender} established this paradigm using RQ-VAE, and several subsequent methods have followed the same residual-based scheme \cite{kong2025minionerec, he2025plum, tang2025reaseq}. These works share a fundamental structural property: codebooks are global across levels and applied to residual representations after subtracting previously selected centroids. As a result, two items that share a code at some level need not be close in the original embedding space. Many extensions remain within this framework, including LETTER \cite{wang2024learnable}, which adds semantic and collaborative regularization to improve RQ-VAE; RVQ \cite{ju2025generative}, which performs multi-stage residual codebook lookup; RQ-means \cite{deng2025onerec}, which replaces the neural autoencoder with K-means-based residual quantization; RRS \cite{zhang2025purely}, which addresses collisions through recursive candidate search; and CoST \cite{zhu2024cost}, HiGR \cite{pang2025higr}, and QINCo2 \cite{vallaeys2025qinco2}, which further augment the residual pipeline with contrastive or conditional reconstruction objectives. While effective, these approaches still optimize quantization first and rely on the induced code indices as identifiers, leaving the hierarchical path structure only indirectly specified. This also gives rise to a large combinatorial code space in which many possible token sequences do not correspond to any real item. In parallel, tree-based methods such as SEATER \cite{si2024generative} and related approaches \cite{hua2023index, wang2024content} move closer to a hierarchical view of indexing. However, these methods typically rely on fixed or otherwise predetermined branching structures, and in some cases introduce additional training objectives that may further entangle representation learning with the indexing process.

Beyond these approaches, several alternative strategies for representing items have been proposed. GPTRec \cite{petrov2023generative} quantizes SVD-based collaborative embeddings, while IDGenRec \cite{tan2024idgenrec} learns textual identifiers directly from metadata. ActionPiece \cite{hou2025actionpiece} adopts a contextual tokenization scheme in which the same item may receive different tokens depending on the interaction context. SETRec \cite{lin2025order} represents items using order-agnostic sets of continuous tokens, and HiD-VAE \cite{fang2025hid} learns hierarchical disentangled identifiers with an explicit uniqueness objective. RPG \cite{hou2025generating} further departs from sequential decoding by predicting long unordered identifiers in parallel.
While these methods represent meaningful advances in generative recommendation, their contributions are tied to specific architectural mechanisms 
that would confound any direct comparison. Benchmarking against them would conflate tokenization quality with model architecture, obscuring the signal we aim to isolate. We therefore focus our evaluation on methods sharing the same autoregressive, discrete-token framework, where tokenization choices can be compared under controlled conditions.

In this paper, we keep the construction simple and directly optimize for the structure that semantic IDs ultimately serve: a path from the root to the target item, where more similar items should share longer prefixes. Based on this perspective, we introduce two key improvements. First, we use adaptive branching, allowing the number of children at each node to vary according to the local data geometry rather than imposing a fixed branching factor. Second, we incorporate mass constraints to prevent prefixes with extremely limited supervision, which is particularly important under the long-tail distributions that characterize recommendation data.

\section{Motivation: Semantic IDs as Tree Paths}
\label{sec:semantic_id_analysis}

In this work, the main research question we address in the context of generative recommendation is the following: \textit{do the discrete tokens in an item's semantic ID carry independent meaning, or is each token's meaning conditioned on the prefix that precedes it?}


As an initial step toward answering this question, we conduct an analysis on the Amazon Reviews dataset and present results on the Beauty dataset as a representative use case. To obtain item semantic IDs, we employ the RQ-VAE tokenizer introduced in TIGER \cite{rajput2023recommender}, as it represents the earliest and most widely adopted tokenization method in generative recommendation to date.
Each item's semantic ID is a 4-token sequence. For the analysis, we define exact-matching masks to measure various mean-similarity patterns between pairs of items. For example, the mask \texttt{1100} selects pairs of items whose semantic IDs share the first two tokens but differ in the last two. Mean similarities are estimated via bootstrap sampling.

Our first analysis uses masks \texttt{1000}, \texttt{1100}, and \texttt{1110} to test whether longer shared prefixes correspond to higher similarity. The results confirm a clear monotonic trend: mean cosine similarities are $0.758$, $0.874$, and $0.920$, respectively, against a global baseline of only $0.554$. Each additional shared prefix token yields a substantial similarity increase. These differences are statistically significant under bootstrap confidence intervals (non-overlapping intervals across conditions), consistent with a coarse-to-fine routing process.

Our second analysis directly compares prefix agreement (\texttt{1000}) against suffix agreement (\texttt{0111}). If token positions were independent, matching three suffix tokens should be highly informative. Instead, sharing only the first token yields a mean cosine similarity of $0.740$, while sharing the last three tokens but differing on the first yields only $0.582$, barely above the global baseline of $0.570$. This gap is statistically significant based on paired bootstrap analysis. Evidently, later tokens do not carry stable global meaning; rather, their interpretation is strongly conditioned on the preceding prefix.


These analyses reveal that semantic IDs behave as paths in a tree rather than tuples of independent attributes. Each token represents a branching decision conditioned on all prior tokens, progressively refining a coarse-to-fine partition. This observation carries a direct methodological implication: since semantic IDs serve as tree paths, it may be preferable to optimize this structure rather than relying on residual quantization objectives and hoping a useful path structure emerges incidentally. This motivates our approach of explicitly constructing semantic IDs as paths in a tree whose branching structure is optimized for the induced semantic organization.

\section{Adaptive Silhouette Tree}
\label{sec:method}
\begin{figure}
    \centering
    \includegraphics[width=0.9\linewidth]{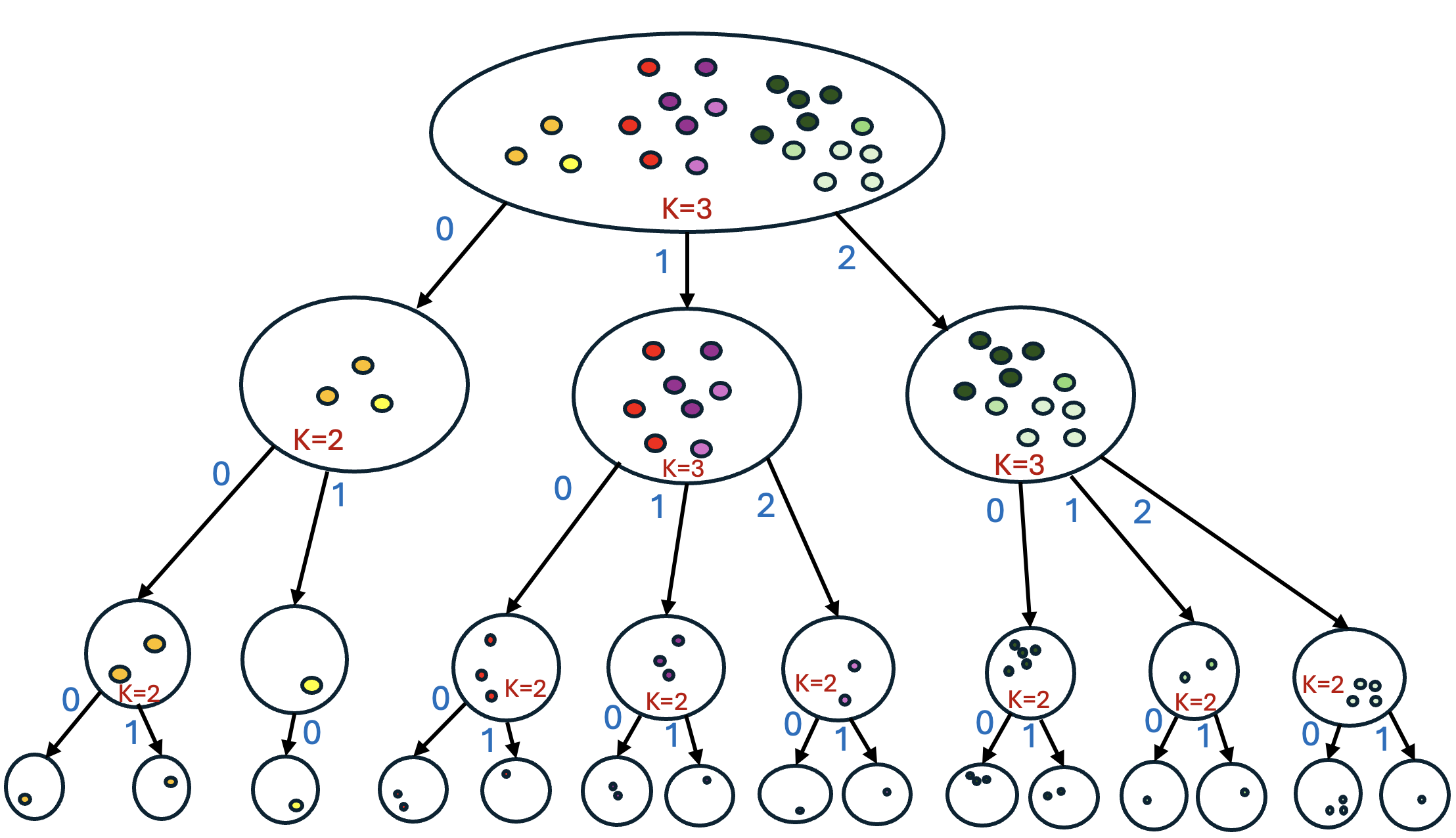}
    \caption{Adaptive Silhouette Tree (AST) method ($L=3$)}
    \label{fig:ast}
\end{figure}

We now describe our approach to constructing semantic IDs for generative
recommendation. We first formalize the problem setup and the structure of
semantic IDs (Section~\ref{sec:setup}). We then introduce the Adaptive
Silhouette Tree (AST)--a top-down divisive clustering procedure that
selects branching factors by maximizing the silhouette coefficient at
each node (Section~\ref{sec:ast}); and discuss its important properties (Section~\ref{sec:ast properties}). We then present
two constrained variants that incorporate item mass information
into the splitting process, ensuring that each branch carries a minimum
fraction of the parent's total mass (Section~\ref{sec:constrained}). 
Finally, we discuss how new items can be incorporated
into an existing tree at inference time (Section~\ref{sec:coldstart}). Throughout, we contrast AST
with RQ-VAE~\cite{lee2022autoregressive}, the most widely adopted
semantic ID method in the generative recommendation literature, to
clarify the former's qualitative advantages.

\subsection{Problem Setup and Semantic IDs}
\label{sec:setup}

Let $\mathcal{I}=\{1,\dots,N\}$ denote the item set, where each item
$i$ is associated with textual metadata~$x_i$. Our goal is to assign
each item a unique semantic ID (SID) of the form
$\mathbf{s}_i = (s_i^{(0)}, \dots, s_i^{(m-1)})$ with
$s_i^{(\ell)} \in \{0,\dots,B{-}1\}$, where $m$ is the code length
and $B$ is the maximum branching factor, or \emph{base}.

We adopt a path-based interpretation: the meaning of token
$s_i^{(\ell)}$ is defined relative to its prefix
$(s_i^{(0)},\dots,s_i^{(\ell-1)})$, not globally. To prevent
ambiguity, we allocate a disjoint token vocabulary
$\mathcal{V}^{(\ell)}=\{\langle c\ell\_0\rangle,\dots,\langle
c\ell\_{B{-}1}\rangle\}$ for each position. Each item's text is
embedded and normalized as
$\tilde{\mathbf{e}}_i = \mathbf{e}_i / \|\mathbf{e}_i\|_2$ with
$\mathbf{e}_i = \mathrm{Enc}(x_i) \in \mathbb{R}^d$. All clustering is performed in this normalized
embedding space.

\subsection{Tree Construction}
\label{sec:ast}



The \emph{Adaptive Silhouette Tree} (AST) generates the first $L = m-1$ tokens of the SID through top-down divisive clustering:

\noindent $\bullet$ \textbf{Node Split:} Starting at the root, each node $v$ containing a subset of items $S_{v} \subseteq \mathcal{I}$ is split into $K_{v}$ children. Every item $i \in S_v$ is assigned a token $s_i^{(\ell)}\in\{0, \dots, K_{v}-1\}$ based on the index of its assigned child node.

\noindent $\bullet$ \textbf{Recursion:} Once the best split is determined for level $\ell$, tokens are assigned and the process repeats independently for each child until $L$ tokens are generated or no feasible split remains.

\noindent $\bullet$ \textbf{Leaf Refinement:} Items sharing the same $L$-token prefix are grouped into buckets. To ensure every SID is unique, the final token $s_i^{(m-1)}$ is assigned based on the item's deterministic rank (e.g., by item ID) within its bucket.

An example of the top-down tree construction is illustrated in Figure~\ref{fig:ast}, assuming $L=3$.

\subsubsection{Adaptive $K_v$ Branching Factors}

For a node $v$ containing $n_v = |S_v|$ items at level~$\ell$, we search over
$K_v \in [K_{\min}(n_v,\ell),\, K_{\max}(n_v)]$. The upper bound
$K_{\max}(n_v) = \min(B,\, n_v{-}1)$ ensures the silhouette score is
defined. The lower bound
  $K_{\min}(n_v,\ell)
  = \max\!\Bigl(2,\;\Bigl\lceil \frac{n_v}{B^{\,L-\ell}} \Bigr\rceil\Bigr)$
enforces a capacity constraint: with $L{-}\ell$ levels remaining, the
subtree can realize at most $B^{L-\ell}$ distinct prefixes, so large
nodes must branch aggressively enough to keep leaf buckets within the
base limit. By induction, this guarantees that every leaf at the final prefix level contains at most~$B$ items, ensuring that suffix
assignment always produces unique SIDs.

\subsubsection{Split Selection}
For each candidate~$K_v$ split size, we run KMeans on the embeddings in the current
node. 
In the default (unconstrained) AST setting, the resulting partition is scored
with the silhouette~\cite{rousseeuw1987silhouettes} coefficient:
  $\mathrm{Sil}(S_v,K_v)
  = \frac{1}{|S_v|} \sum_{i \in S_v}
    \frac{b_i - a_i}{\max(a_i,\, b_i)}$,
where $a_i$ is the mean intra-cluster distance and $b_i$ the mean
nearest-cluster distance. Candidate split values $K_v$ are evaluated in ascending order;
we select the feasible split with the highest score, breaking
ties in favor of smaller values, 
and apply early stopping when the score
fails to improve over several consecutive candidates. For large nodes,
the silhouette can be estimated from a random subsample.
\subsection{AST Properties}
\label{sec:ast properties}
\subsubsection{Structural Properties}
\label{sec:ast:structure}

Let $S_p$ denote the subset of items in $\mathcal{I}$ that share the same (SID) prefix $p$. Because the tree is built by recursive partitioning, the item sets
associated with all prefixes form a \textit{laminar family}: for any two
prefixes $p,p'$, either $S_p \cap S_{p'} = \emptyset$ or one contains
the other. Each prefix group is exactly one KMeans cluster obtained by
partitioning a single parent in the original normalized embedding
space, so sharing a prefix directly implies geometric proximity. In
RQ-VAE, by contrast, level-$\ell$ codebooks are shared globally and
applied to residuals after subtracting all previous centroids; two
items with different level-0 codes but identical level-1 codes may have
residuals near the same codebook vector while their original embeddings
are arbitrarily far apart.

The laminar structure also ensures full code utilization: every emitted
code corresponds to at least one item, since KMeans is run only on
items present in each node and empty clusters are discarded. RQ-VAE is
well known to suffer from codebook collapse, where a significant
fraction of entries receive no assignments. A related advantage
concerns decodability. The set of valid SID sequences in AST is exactly
the set assigned to real items, whereas RQ-VAE defines codes via a
Cartesian product of $m$ codebooks, yielding $B^m$ possible sequences
of which only~$N$ are valid. The fraction of hallucinated sequences, $1 - N/B^m$, may even exceed 99.9\%, wasting beam-search capacity on invalid paths.

\subsubsection{Clustering Distortion and the Benefit of Adaptive Branching}

Although AST ultimately assigns discrete semantic IDs, its construction is not based on residual quantization in the style of RQ-VAE. Instead, AST builds a hierarchical clustering tree directly in the original normalized embedding space. For this reason, it is more precise to discuss \emph{clustering distortion} rather than quantization error.

Let $\mu_{\mathrm{leaf}(i)}$ denote the centroid of the final leaf cluster containing item $i$. The total clustering distortion of the tree is
$\sum_i \left\lVert \tilde{e}_i - \mu_{\mathrm{leaf}(i)} \right\rVert^2$.
Because AST is constructed through recursive partitioning, this quantity admits a telescoping decomposition:
$\sum_i \left\lVert \tilde{e}_i - \mu_{\mathrm{leaf}(i)} \right\rVert^2
=
\sum_{\ell} \sum_{v \text{ at level } \ell} \Delta_v$,
where each $\Delta_v \ge 0$ is the reduction in within-cluster variance obtained by splitting node $v$. This decomposition is useful because every term is defined in the same original embedding space. As a result, one can directly interpret which splits meaningfully improve the semantic organization of the tree and which contribute little.

This is an important distinction from residual quantization methods such as RQ-VAE. In RQ-VAE, later assignments are made in residual spaces that depend on previous code selections, so the meaning of a level is harder to interpret geometrically in the original space. In AST, by contrast, each split corresponds to an explicit partition of the parent cluster in the same embedding space, making the hierarchy easier to analyze and diagnose.

Adaptive branching further improves this objective. For a fixed total number of leaf nodes $L_{\mathrm{total}}$, let $W^*_{\mathrm{fixed}}$ denote the minimum achievable within-cluster variance under a tree with a constant branching factor, and $W^*_{\mathrm{adaptive}}$ denote corresponding minimum under a tree whose branching factor may vary from node to node, subject to $\forall{v}: 2 \le K_v \le B$. Then
$W^*_{\mathrm{adaptive}} \le W^*_{\mathrm{fixed}}$.
The reason is that a fixed-branching tree is a special case of an adaptive tree. Moreover, when the item distribution is non-uniform, adaptive branching is strictly more expressive: dense and heterogeneous regions can be partitioned more finely, while sparse or homogeneous regions remain coarse. By selecting the branching factor locally based on silhouette quality, AST effectively allocates modeling capacity. 

Overall, this perspective clarifies that AST should be understood as a hierarchical clustering procedure that produces semantic ID paths, rather than as a residual quantization method. The resulting distortion objective is therefore directly tied to the quality of the tree partition itself.

\subsection{Constrained Split Variants}
\label{sec:constrained}
In practice, item interactions follow a highly skewed distribution, where a small fraction of items accounts for the majority of user engagement. Unconstrained clustering is agnostic to this popularity structure and can therefore allocate large portions of the interaction mass to a few branches while leaving others sparsely populated. As a result, certain prefix paths receive very limited supervision during training. These under-supported paths are difficult for the generative model to learn reliably, leading to poor estimation of conditional probabilities and degraded generation quality when such prefixes are encountered at inference time.
Referring back to the example in Figure~\ref{fig:ast}; let us assume that each child node must inherit at least one-third of its parent's mass; consequently, the tree's root must now have two children instead of three. 

We next present two constrained variants, \textbf{AST-V1} (Section~\ref{sec:optionA}) and \textbf{AST-V2} (Section~\ref{sec:optionB}), that address this limitation by incorporating item mass information directly into the splitting process. Both variants ensure that every child branch carries at least a minimum fraction of the parent's total interaction mass, yielding a tree whose structure is better aligned with the downstream recommendation objective. The two variants occupy different points on the tradeoff between simplicity and fidelity to the constrained objective. AST-V1 is a lightweight correction that retains standard KMeans centroids and enforces feasibility only when needed, scoring splits with the unweighted silhouette. AST-V2 incorporates mass throughout, unifying scoring, centroid computation, and the balance constraint into a single coherent objective. We compare both variants empirically in Section~\ref{sec:experiments}.
\subsubsection{Incorporating mass information}
As an optional requirement, we now associate each item~$i$ with a nonnegative mass~$p_i$
(e.g., interaction frequency), normalized so that $\sum_i p_i = 1$.
Within a node~$S_v$, local masses are
$\pi_i = p_i \,/\, \sum_{r \in S_v} p_r; \forall i\in{S_v}$. For a given $\rho\in(0,1]$, both constrained variants
enforce that every child~$C_j$ of a split satisfy
$\sum_{i \in C_j} \pi_i \geq \rho$, ensuring each branch carries at
least a fraction~$\rho$ of the parent's mass. This implies
$K_v \leq \lfloor 1/\rho \rfloor$, tightening the candidate range to
$K_v \;\in\;
  \Bigl[\,
    \max\!\bigl(K_{\min}(n_v,\ell),\, 2\bigr),\;
    \min\!\bigl(K_{\max}(n_v),\, \lfloor 1/\rho \rfloor\bigr)
  \,\Bigr]$.
If this interval is empty, no feasible split exists and the recursion halts, signaling that the items cannot be further separated without violating capacity or mass requirements. RQ-VAE offers no analogous mechanism for controlling how item mass is distributed across codebook entries. When masses are unavailable, we default to uniform weights.

\subsubsection{AST-V1: Fixed-Center Feasibility Correction}
\label{sec:optionA}

AST-V1 treats KMeans as a semantic initializer and intervenes only
when its assignments violate the mass constraint. For a given node $v$ and candidate partition size~$K_v$, we
run standard KMeans and obtain centers
$\boldsymbol{\mu}_1, \dots, \boldsymbol{\mu}_{K_{v}}$. If the partition is
already feasible, we keep it. Otherwise, we hold the centers fixed and
solve the reassignment problem:
\vspace{-0.2in}
\begin{align}
  \min_{\{z_{ij}\}} \quad
    & \sum_{i \in S_v} \sum_{j=1}^{K_v}
      \|\tilde{\mathbf{e}}_i - \boldsymbol{\mu}_j\|_2^2 \; z_{ij}
      \label{eq:milp-obj} \\
  \text{s.t.} \quad
    & \sum_{j=1}^{K_v} z_{ij} = 1,
      \qquad \forall\, i \in S_v,
      \label{eq:milp-assign} \\
    & \sum_{i \in S_v} \pi_i \, z_{ij} \geq \rho,
      \qquad \forall\, j \in \{1,\dots,K_{v}\},
      \label{eq:milp-mass} \\
    & z_{ij} \in \{0,1\}.
      \label{eq:milp-binary}
\end{align}
This mixed-integer linear program (MILP) finds the globally optimal feasible assignment given the fixed centers when solved exactly (i.e., without approximation). The split is then scored with the standard silhouette $\mathrm{Sil}(S_v,K_v)$.

Denoting the unconstrained KMeans distortion by $W_{\mathrm{KM}}$ and
the corrected distortion by~$W_{\mathrm{A}}$, we have
$W_{\mathrm{A}} \geq W_{\mathrm{KM}}$ since feasibility restricts the
assignment space, but $W_{\mathrm{A}} \leq W(\mathbf{z})$ for any other
feasible assignment~$\mathbf{z}$ to those centers. AST-V1, therefore,
incurs the minimal distortion penalty necessary to achieve feasibility
while preserving the original cluster geometry.

\subsubsection{AST-V2: Mass-Aware Constrained Refinement}
\label{sec:optionB}

AST-V2 integrates mass into both scoring and centroid computation
rather than treating it as a post-hoc correction. Splits are scored
with the weighted silhouette
  $\mathrm{WSil}(S_v,K_v) = \sum_{i \in S_v} \pi_i \, \mathrm{Sil}(\{i\},K_v)$,
giving greater influence to high-mass items and reducing to the
mean silhouette under uniform masses. When estimated from a
subsample, drawing items in proportion to~$\pi_i$ yields an unbiased
estimator, whereas uniform subsampling would bias toward low-mass
items.

Starting from KMeans centers, AST-V2 performs a short alternating
refinement. Step~(i) solves the same MILP as
AST-V1 (Eq.~\eqref{eq:milp-obj}--Eq.\eqref{eq:milp-binary}), yielding the
globally optimal feasible assignment for the current centers.
Step~(ii) recomputes each center as the mass-weighted mean:
$\boldsymbol{\mu}_j \leftarrow \sum_{i:\, z_{ij}=1} \pi_i\,\tilde{\mathbf{e}}_i \,/\, \sum_{i:\, z_{ij}=1} \pi_i$.
These two steps alternate for a small fixed number of rounds or until
convergence. Let $W^{(t)}$ denote the total distortion after
iteration~$t$. Step~(i) cannot increase distortion because it solves a
global optimum over feasible assignments; Step~(ii) cannot increase it
because the weighted mean uniquely minimizes squared distances for
fixed assignments. Hence
$W^{(0)} \geq W^{(1)} \geq \cdots \geq 0$, providing the constrained
analogue of Lloyd's convergence guarantee~\cite{lloyd1982least}, strengthened by exact
optimization in the assignment step. The quantity minimized by the full procedure is
  $W_\pi
  = \sum_{j=1}^{K_v} \sum_{i:\, z_{ij}=1}
    \pi_i \, \|\tilde{\mathbf{e}}_i - \boldsymbol{\mu}_j\|^2$,
which assigns lower reconstruction error to frequently interacted
items. This is a desirable inductive bias: items appearing more often
in training sequences receive more faithful representations, aligning indexing with the downstream loss. The telescoping
variance decomposition extends directly as
$W_\pi = \sum_\ell \sum_v \Delta_v^{(\pi)}$, where each
$\Delta_v^{(\pi)} \geq 0$ is the mass-weighted variance reduction at
node~$v$.

\subsection{Cold-Start Assignment}
\label{sec:coldstart}

Given a new item, AST assigns a semantic ID by encoding its text and greedily traversing the tree using nearest-centroid decisions at each level. This requires $L = m{-}1$ steps with at most~$B$ comparisons per level, yielding $O(BdL)$ cost independent of catalog size~$N$. The new item inherits the same hierarchical nesting and geometric coherence as items present during construction.

In RQ-VAE, assignment has comparable $O(BdL)$ cost, but each step operates on a residual depending on all previous codes, so greedy errors at one level compound downstream with no guarantee of semantic coherence in the original space. More broadly, cold start remains partly a model-side issue: even with a valid SID, the downstream recommender has not been trained to produce that new token sequence.

\begin{table*}[t]
\centering
\caption{Performance comparison across all domains. Values are reported as percentages. Best performing method is \textbf{bolded}, second-best is \underline{underlined}. $^*$ indicates statistical significance at $p < 0.05$ compared to the strongest non-AST baseline.}
\label{tab:main_results}
\resizebox{\textwidth}{!}{%
\begin{tabular}{l|l|cccccccccc|cc}
\hline
Domain & Metric & RQ-VAE & RQ-VAE-L & Qinco2 & RVQ & CoST & RQ-Means & RQ-Means-B & RRS & SEATER & CRVQAE & AST-V1 & AST-V2 \\
\hline
\multirow{4}{*}{Beauty}
& R@5  & 4.15 & 4.31 & 4.61 & 4.48 & \underline{4.63} & 4.51 & \underline{4.63} & 4.27 & 4.50 & 4.40 & \textbf{5.12} & \textbf{5.12}$^*$ \\
& R@10 & 6.09 & 6.18 & 6.91 & 6.77 & 6.61 & 6.98 & 6.33 & 6.44 & 6.65 & 6.64 & \underline{7.31} & \textbf{7.51}$^*$ \\
& N@5  & 2.93 & 2.86 & 3.08 & 2.93 & 3.12 & 3.11 & 2.89 & 2.96 & 3.00 & 2.93 & \underline{3.50} & \textbf{3.51}$^*$ \\
& N@10 & 3.56 & 3.47 & 3.82 & 3.67 & 3.76 & 3.86 & 3.55 & 3.66 & 3.69 & 3.65 & \underline{4.20} & \textbf{4.28}$^*$ \\
\hline
\multirow{4}{*}{Toys}
& R@5  & 3.98 & 3.86 & 3.98 & 4.63 & 4.31 & 4.45 & 3.96 & 3.92 & 4.44 & 4.35 & \underline{5.16} & \textbf{5.27}$^*$ \\
& R@10 & 6.01 & 5.86 & 6.01 & 7.07 & 6.42 & 6.67 & 5.91 & 6.04 & 6.44 & 6.58 & \underline{7.75} & \textbf{7.94}$^*$ \\
& N@5  & 2.57 & 2.58 & 2.57 & 2.93 & 2.90 & 2.80 & 2.65 & 2.59 & 3.07 & 2.92 & \underline{3.41} & \textbf{3.45}$^*$ \\
& N@10 & 3.23 & 3.23 & 3.23 & 3.72 & 3.58 & 3.52 & 3.28 & 3.27 & 3.72 & 3.64 & \underline{4.24} & \textbf{4.31}$^*$ \\
\hline
\multirow{4}{*}{Sports}
& R@5  & 2.41 & 2.46 & 2.73 & 2.73 & 2.62 & 2.64 & 2.37 & 2.52 & 2.34 & 2.67 & \underline{3.01} & \textbf{3.09}$^*$ \\
& R@10 & 3.48 & 3.67 & 4.08 & \underline{4.09} & 4.02 & 4.08 & 3.66 & 3.87 & 3.73 & 4.01 & \textbf{4.62} & \textbf{4.62}$^*$ \\
& N@5  & 1.58 & 1.64 & 1.71 & 1.79 & 1.72 & 1.72 & 1.53 & 1.66 & 1.50 & 1.72 & \underline{2.02} & \textbf{2.03}$^*$ \\
& N@10 & 1.93 & 2.03 & 2.15 & 2.23 & 2.17 & 2.18 & 1.94 & 2.10 & 1.95 & 2.16 & \textbf{2.54}$^*$ & \underline{2.51} \\
\hline
\multirow{4}{*}{Health}
& R@5  & 3.97 & 3.96 & 4.13 & 4.12 & \underline{4.33} & 3.95 & 3.82 & 4.13 & 3.60 & 4.15 & 4.32 & \textbf{4.48}$^*$ \\
& R@10 & 5.85 & 5.92 & 6.06 & 5.86 & 6.30 & 5.70 & 5.47 & 6.07 & 5.39 & 5.99 & \underline{6.40} & \textbf{6.64}$^*$ \\
& N@5  & 2.64 & 2.66 & 2.69 & 2.68 & 2.86 & 2.59 & 2.49 & 2.65 & 2.34 & 2.69 & \underline{2.89} & \textbf{2.98}$^*$ \\
& N@10 & 3.25 & 3.29 & 3.31 & 3.24 & 3.49 & 3.16 & 3.02 & 3.28 & 2.91 & 3.28 & \underline{3.58} & \textbf{3.68}$^*$ \\
\hline
\multirow{4}{*}{Home}
& R@5  & 1.51 & 1.67 & 1.69 & 1.61 & \underline{1.86} & 1.62 & 1.40 & 1.44 & 1.54 & 1.77 & 1.68 & \textbf{2.06}$^*$ \\
& R@10 & 2.28 & 2.49 & 2.59 & 2.40 & 2.63 & 2.46 & 2.13 & 2.25 & 2.26 & 2.61 & \underline{2.64} & \textbf{3.02}$^*$ \\
& N@5  & 1.03 & 1.12 & 1.12 & 1.06 & \underline{1.27} & 1.03 & 0.89 & 0.94 & 1.01 & 1.19 & 1.13 & \textbf{1.36}$^*$ \\
& N@10 & 1.28 & 1.38 & 1.41 & 1.32 & \underline{1.51} & 1.31 & 1.12 & 1.20 & 1.24 & 1.46 & 1.43 & \textbf{1.67}$^*$ \\
\hline
\end{tabular}%
}
\end{table*}

\section{Experiments}\label{sec:experiments}

\begin{table}[t]
\centering
\caption{Dataset statistics.}
\label{tab:dataset_stats}
\begin{tabular}{l|cccc}
\hline
Dataset & \#Users & \#Items & Sparsity $(\%)$ & Avg Length \\
\hline
Beauty & 22,363 & 12,101 & 99.93\% & 8.88  \\
Sports & 35,598 & 18,357 & 99.95\% & 8.32  \\
Toys & 19,412 & 11,924  & 99.93\% & 8.63  \\
Health & 38,609 & 18,534 & 99.95\% & 8.97  \\
Home & 66,519 & 28,237 & 99.97\% & 8.29 \\
\hline
\end{tabular}
\end{table}

\subsection{Experimental Setup}



\paragraph{Datasets.}
We evaluate our method on five Amazon Review domains~\cite{he2016ups}: \textit{Beauty}, \textit{Toys and Games}, \textit{Sports and Outdoors}, \textit{Health and Personal Care}, and \textit{Home and Kitchen}, which span diverse product categories and catalog sizes to test robustness across recommendation settings. All datasets undergo 5-core filtering, and interaction sequences are truncated to a maximum length of 20. Item representations are obtained by concatenating each product's \textit{title}, \textit{categories}, and \textit{description} fields into a single
text fed to the item encoder.

\paragraph{Baselines.}
We compare AST against a diverse set of semantic ID tokenization methods used in generative recommendation, spanning residual quantization, clustering-based tokenization, tree-based methods, and explicit collision-handling approaches. We further specify in parentheses notable generative recommendation models that utilize each method. 

\begin{itemize}[leftmargin=*,nosep,topsep=2pt]

\item \textbf{RQ-VAE}~\cite{lee2022autoregressive} (TIGER~\cite{rajput2023recommender},  MGR~\cite{zhu2025beyond}, Glass~\cite{cao2026glass}) generates semantic IDs through multi-level residual quantization within a variational autoencoder framework. It is one of the most widely adopted tokenizers in generative recommendation.

\item \textbf{RQ-VAE-L} (LETTER~\cite{wang2024learnable}) extends RQ-VAE by introducing diversity regularization that encourages fuller codebook utilization, thereby improving the quality of the learned semantic IDs.

\item \textbf{CRVQAE} (HiGR~\cite{pang2025higr}) augments the RQ-VAE framework with a contrastive residual quantization objective to improve the discriminability of discrete semantic representations.

\item \textbf{Qinco2}~\cite{vallaeys2025qinco2} is a neural residual quantization method in which each quantization step is conditioned on the reconstruction induced by all previously selected codes, yielding more expressive codeword assignments.

\item \textbf{RVQ}~\cite{esser2021taming} (GRID~\cite{ju2025generative}) is a standard residual vector quantization method that represents each item as a sequence of codewords obtained by recursively quantizing residual errors.

\item \textbf{CoST}~\cite{zhu2024cost} replaces the reconstruction objective with contrastive supervision, learning semantic IDs that better preserve semantic similarity and relational structure in discrete token space.

\item \textbf{RQ-Means} (GPR~\cite{zhang2026gpr}, sJSR~\cite{penha2025sids}) is a clustering-based residual quantization approach that constructs semantic IDs in a coarse-to-fine manner by applying $K$-means to successive residual vectors.

\item \textbf{RQ-Means-B} (EAGER~\cite{wang2024eager}, OneRec~\cite{deng2025onerec}, GRID~\cite{ju2025generative}) is a balanced variant of RQ-Means that enforces a more uniform token distribution across the codebook, mitigating codebook collapse.

\item \textbf{RRS}~\cite{zhang2025purely} addresses semantic ID collisions by recursively searching over residual-based candidates until a unique code assignment is found for each item.

\item \textbf{SEATER}~\cite{si2024generative} is a tree-based method that constructs semantic IDs via hierarchical tokenization and refines them with contrastive supervision to capture multi-granularity item semantics.

\end{itemize}

\paragraph{Implementation Details.}
For fair comparison, all tokenization methods share the same \texttt{Qwen3-Embedding-0.6B}~\cite{zhang2025qwen3} item embedding backbone. Baseline tokenizers use configurations from their original papers where applicable; otherwise, we tune the number of layers and base size per dataset. All tokenization strategies feed into the same generative recommender to isolate the tokenization effect: a T5-based encoder--decoder~\cite{raffel2020exploring} with 6 encoder and 6 decoder layers, trained with dropout $0.15$, initializer factor $0.02$, layer norm epsilon $10^{-6}$, batch size $256$, learning rate $10^{-3}$, weight decay $10^{-4}$, bfloat16 precision, and early stopping with patience $5$ (retaining only the best checkpoint). Our tokenization employs adaptive hierarchical clustering using KMeans~\cite{hartigan1979algorithm} (up to $50$ iterations, $10$ random initializations). At each internal node, the branching factor is selected via silhouette score under Euclidean distance with patience $5$ over candidate values. For large clusters, silhouette scores are computed on a subsample of up to $2{,}000$ items. All models are implemented with \texttt{Transformers}~\cite{wolf2020transformers} and \texttt{scikit-learn}~\cite{pedregosa2011scikit}, and all experiments run on a single NVIDIA RTX 6000 GPU.

\paragraph{Evaluation Metrics.}
We report Recall@$k$ and NDCG@$k$ for $k \in \{5,10\}$ under a leave-last-out protocol: the last interaction per user is used for testing, the second-to-last for validation, and the rest for training~\cite{rajput2023recommender, wang2024content, ju2025generative}.

\subsection{Main Results}

Table~\ref{tab:main_results} compares our AST variants against the full set of semantic ID tokenization baselines. The results reveal a clear and consistent advantage for our approach. Specifically, AST-V2 achieves the best average performance across all four metrics, reaching \textbf{4.00} R@5, \textbf{5.95} R@10, \textbf{2.67} N@5, and \textbf{3.29} N@10. Relative to the strongest non-AST baseline, this corresponds to average improvements of \textbf{12.68\%} on R@5, \textbf{13.55\%} on R@10, \textbf{12.66\%} on N@5, and \textbf{13.45\%} on N@10, yielding an overall relative gain of \textbf{13.08\%} across all datasets and metrics. Moreover, AST-V2 ranks best or tied-best in every dataset–metric combination, confirming that the improvements are not confined to a single domain or evaluation measure. These results strongly support our central claim: semantic IDs are better modeled as hierarchical paths rather than independent token sets, and constructing them via an adaptive, balanced tree yields substantially stronger generative recommendation performance.

\subsection{Performance on Head and Tail Items}

We evaluate AST on \emph{head} and \emph{tail} items separately to understand behavior across the popularity distribution. Following standard practice \cite{adamopoulos2014over, yin2012challenging, sreepada2020mitigating}, items are sorted by popularity; the smallest prefix covering roughly 80\% of interactions forms the \emph{head}, and the rest forms the \emph{tail}.

\begin{figure}[t]
    \centering
    \begin{subfigure}[t]{0.48\linewidth}
        \centering
        \includegraphics[width=\linewidth]{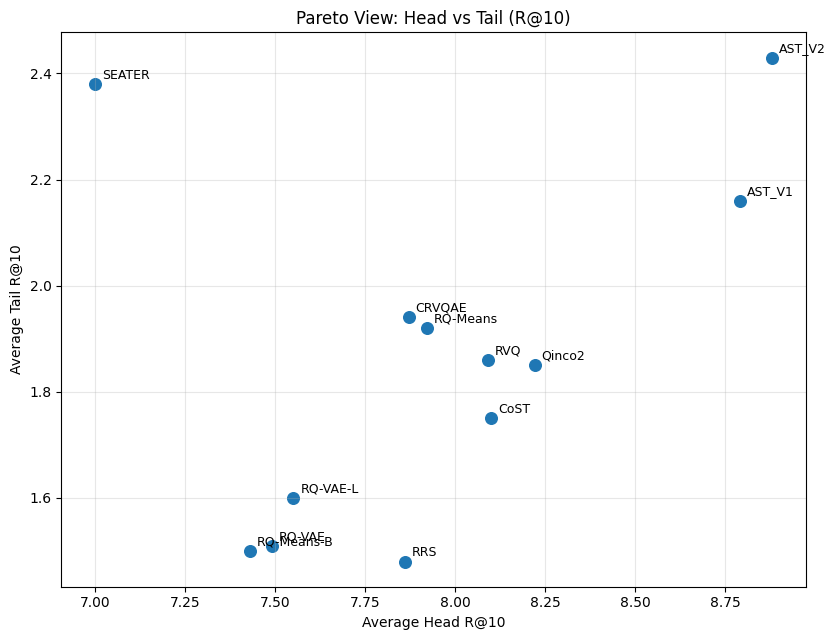}
        \caption{Pareto view of average Head R@10 versus average Tail R@10. Methods on the frontier are not dominated on both axes simultaneously.}
        \label{fig:pareto_r10}
    \end{subfigure}
    \hfill
    \begin{subfigure}[t]{0.48\linewidth}
        \centering
        \includegraphics[width=\linewidth]{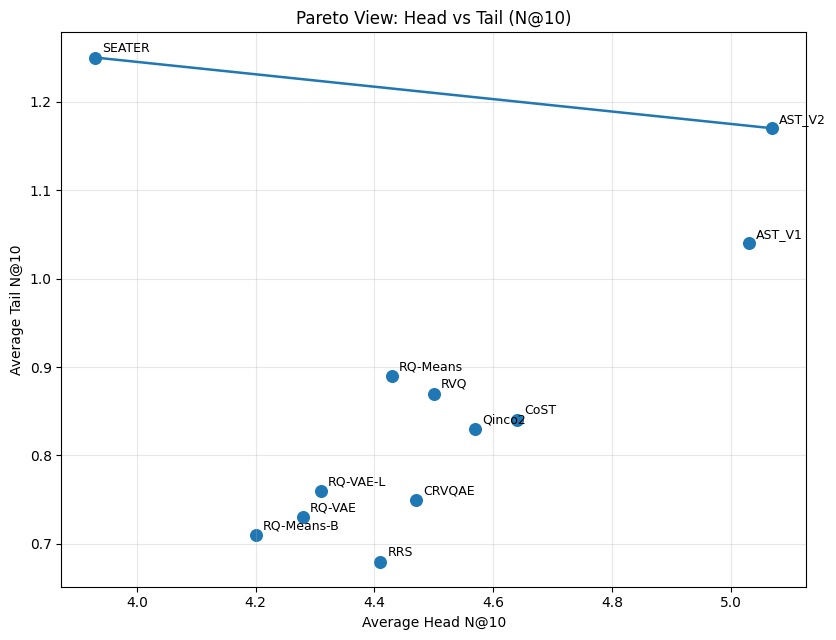}
        \caption{Pareto view of average Head N@10 versus average Tail N@10. AST-V2 remains on the frontier.}
        \label{fig:pareto_n10}
    \end{subfigure}
    \caption{Pareto front analysis of head versus tail performance for Recall@10 (left) and NDCG@10 (right).}
    \label{fig:pareto_combined}
\end{figure}

Overall, \textbf{AST-V2 obtains the best average Head R@10 and Head N@10}, while simultaneously achieving \textbf{the best average Tail R@10 and the second-best average Tail N@10}. Thus, AST does not gain on one side of the spectrum at the expense of the other.
To illustrate these facts clearly, Figures~\ref{fig:pareto_r10} and~\ref{fig:pareto_n10} present Pareto-style plots of average head versus tail performance, where methods closer to the upper-right corner are preferable. In both plots, \textbf{AST-V2 lies on the Pareto frontier}, confirming one of the strongest head/tail trade-offs across methods. Some baselines are skewed: SEATER is strong on the tail but weaker on the head, while others show the opposite pattern.

We further quantify directional bias with two orientation measures. The \textbf{rank-based orientation} captures whether a method's relative ranking shifts between head and tail. Let $r_H^j(m)$ and $r_T^j(m)$ be the rank of method $m$ under metric $j$ on head and tail items. We define:
$\mathrm{Orient}_{\mathrm{rank}}(m)
=
\frac{1}{|k|}
\sum_{j \in k}
\frac{r_T^j(m)-r_H^j(m)}{M-1}$.
Positive values indicate head orientation, negative values tail orientation, and near-zero values balanced behavior. The \textbf{score-based orientation} instead uses normalized metric values. Head and tail scores $s_H(m,j)$, $s_T(m,j)$ are each divided by the median across methods to ensure comparable scales:
$u_H(m,j)=\frac{s_H(m,j)}{\mathrm{median}_{m'}\, s_H(m',j)}$ and
$u_T(m,j)=\frac{s_T(m,j)}{\mathrm{median}_{m'}\, s_T(m',j)}$.
The orientation is then averaged over metrics:
$\mathrm{Orient}_{\mathrm{score}}(m)
=
\frac{1}{|k|}
\sum_{j \in k}
\frac{u_H(m,j)-u_T(m,j)}{u_H(m,j)+u_T(m,j)}$.

\begin{figure}[t]
    \centering
    \includegraphics[width=0.9\linewidth]{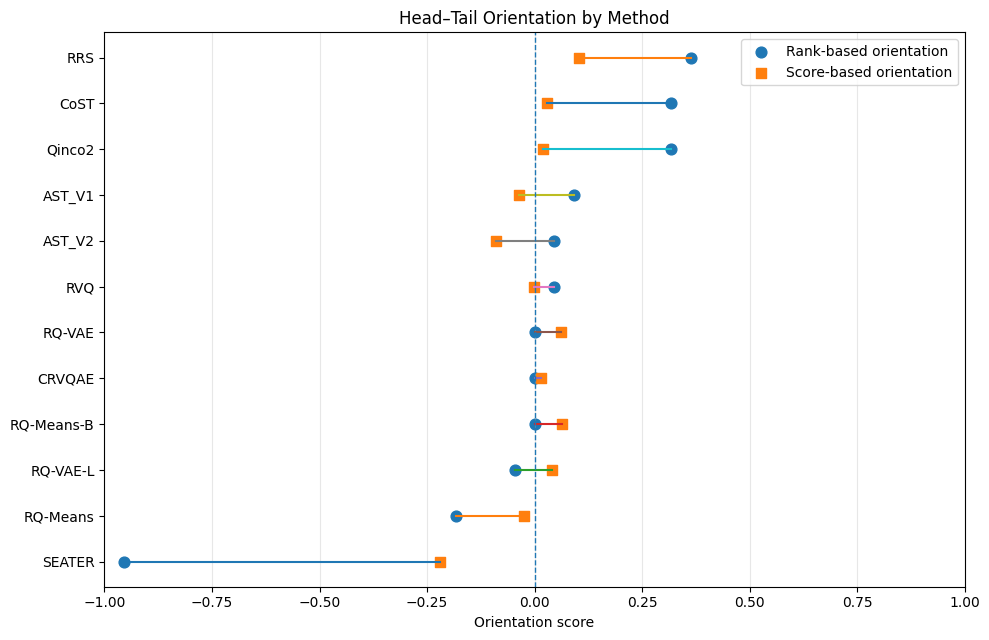}
    \caption{Rank-based and score-based orientation. Positive values indicate head orientation, negative values tail orientation, and near-zero values balanced behavior.}
    \label{fig:orientation_scores}
\end{figure}

Figure~\ref{fig:orientation_scores} compares both views. Some baselines are clearly biased: SEATER is strongly tail-oriented, while RRS is noticeably head-oriented. In contrast, \textbf{AST remains close to the center under both definitions}. AST-V2 has a near-zero rank-based score and only a mildly negative score-based score, indicating largely balanced behavior with a slight tail tendency.

Overall, \textbf{AST is well balanced across the popularity spectrum.} It achieves the strongest head performance while maintaining top-tier tail quality, without the ranking instability observed in several baselines. This makes AST substantially more attractive than methods that improve one regime only at the cost of the other.

\subsection{Sensitivity Analysis}

We study how the probability mass constraints, semantic ID length, and embedding backbone affect performance.

\begin{figure*}[t]
    \centering
    \begin{subfigure}[b]{0.19\textwidth}
        \centering
        \includegraphics[width=\linewidth]{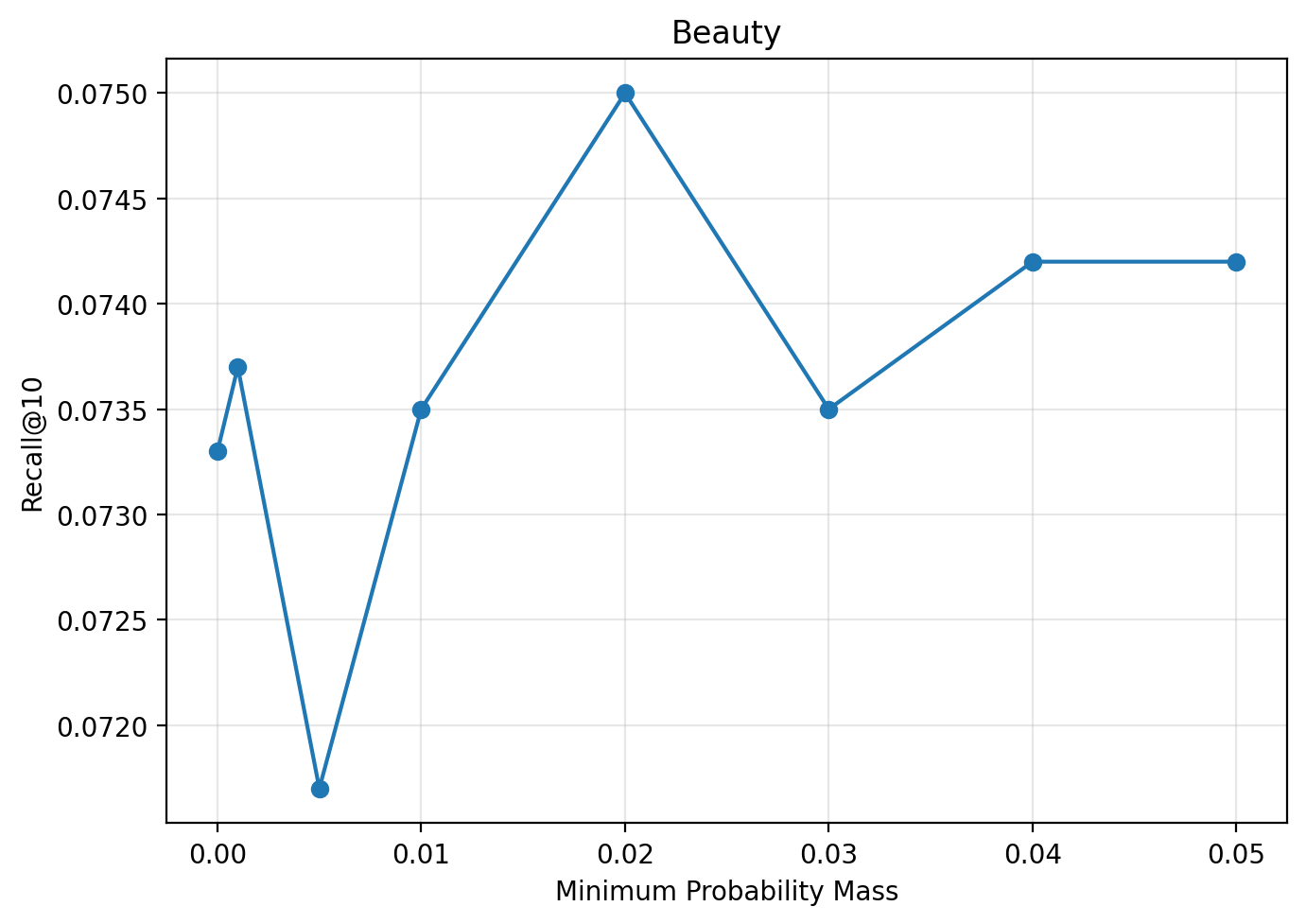}
        \caption{Beauty}
        \label{fig:mass_beauty}
    \end{subfigure}
    \hfill
    \begin{subfigure}[b]{0.19\textwidth}
        \centering
        \includegraphics[width=\linewidth]{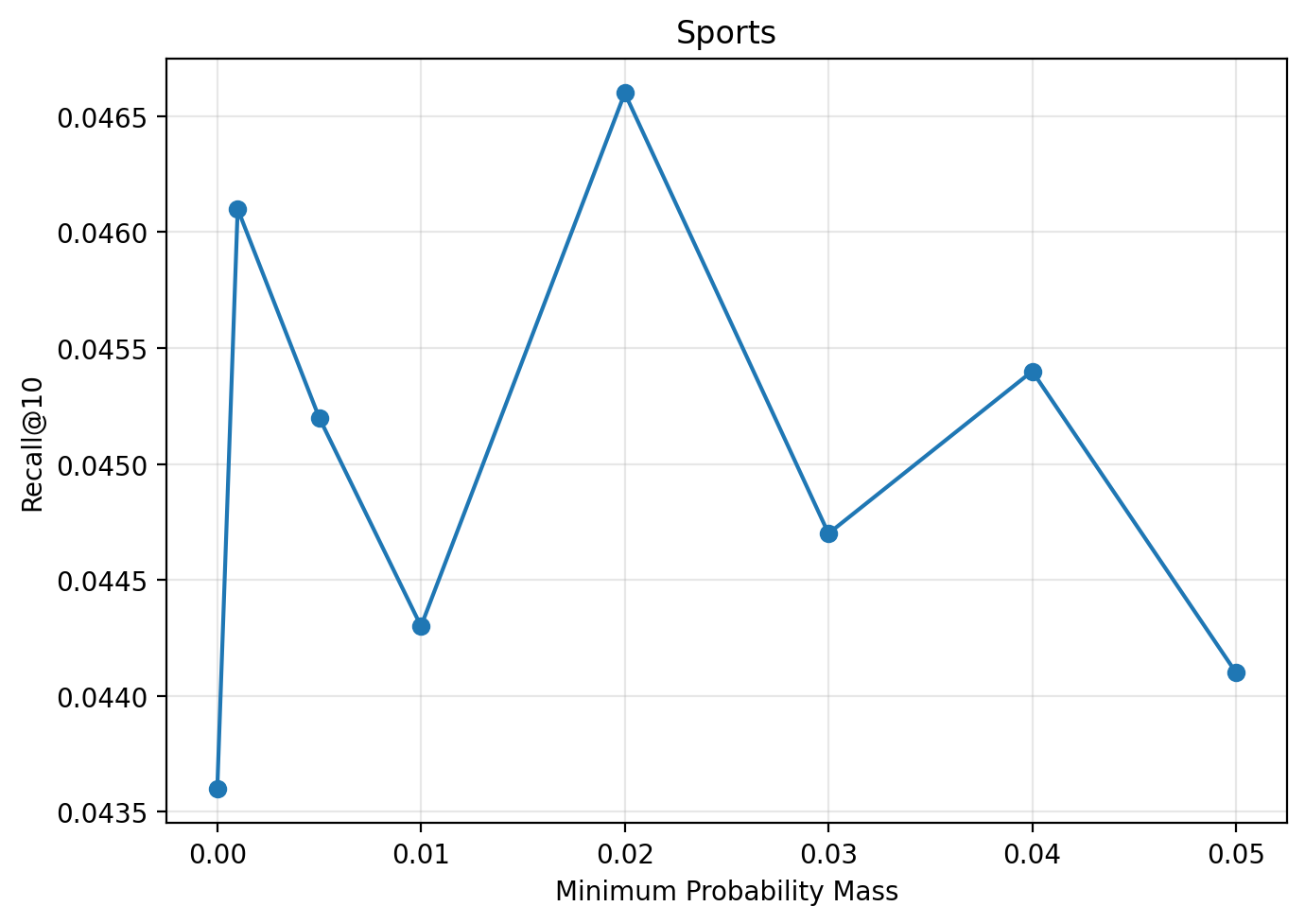}
        \caption{Sports}
        \label{fig:mass_sports}
    \end{subfigure}
    \hfill
    \begin{subfigure}[b]{0.19\textwidth}
        \centering
        \includegraphics[width=\linewidth]{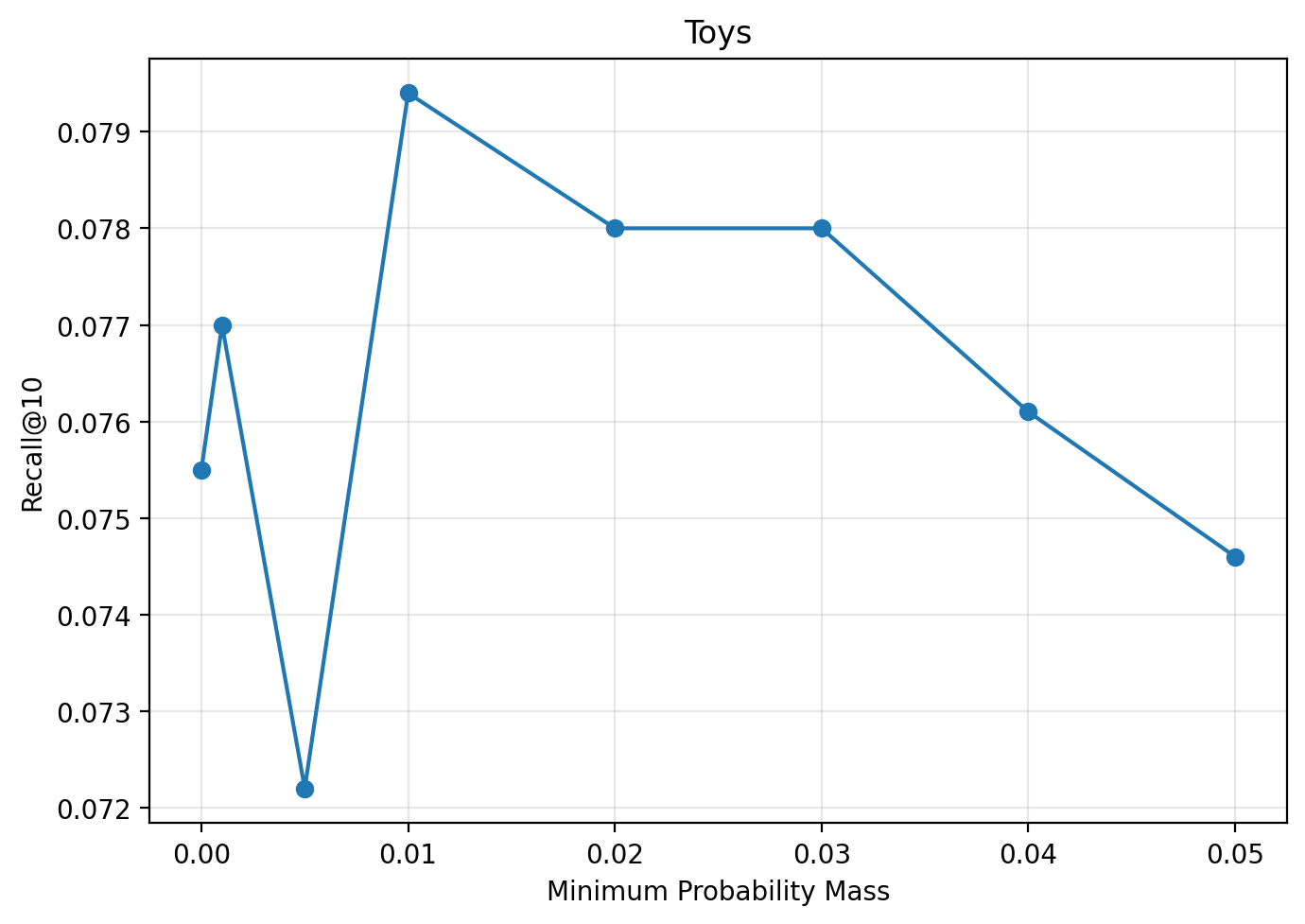}
        \caption{Toys}
        \label{fig:mass_toys}
    \end{subfigure}
    \hfill
    \begin{subfigure}[b]{0.19\textwidth}
        \centering
        \includegraphics[width=\linewidth]{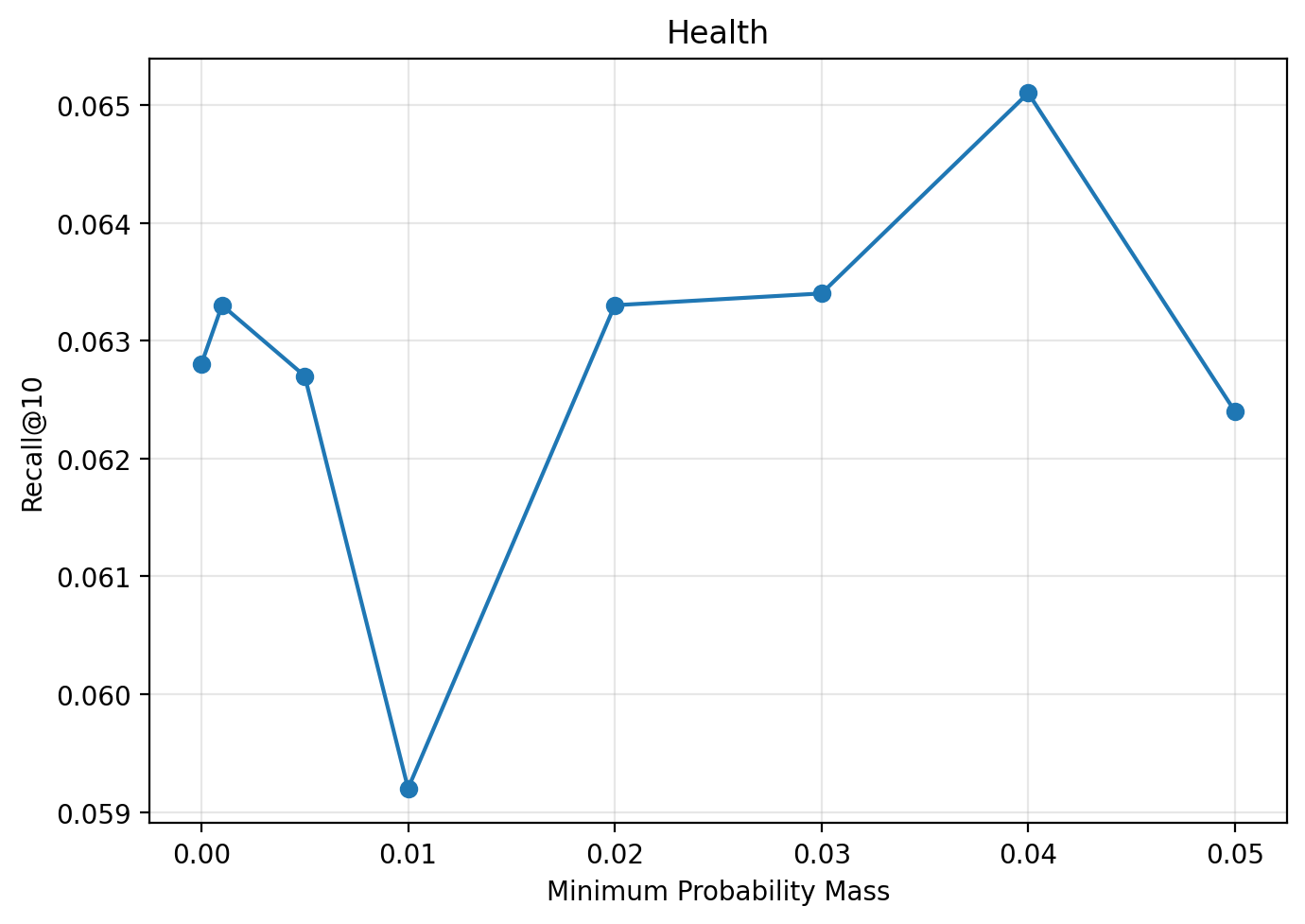}
        \caption{Health}
        \label{fig:mass_health}
    \end{subfigure}
    \hfill
    \begin{subfigure}[b]{0.19\textwidth}
        \centering
        \includegraphics[width=\linewidth]{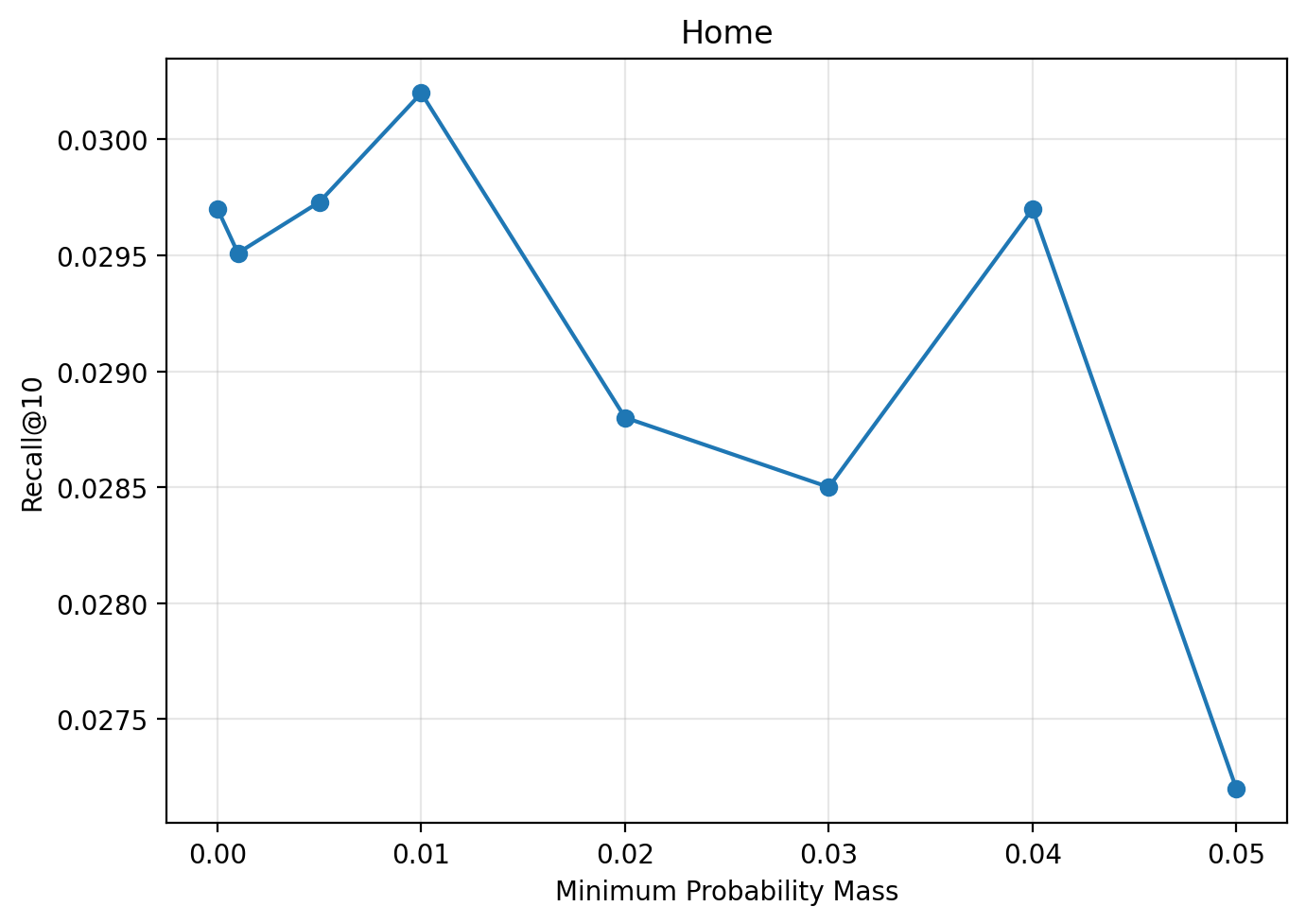}
        \caption{Home}
        \label{fig:mass_home}
    \end{subfigure}
    \caption{Recall@10 as a function of the minimum relative probability mass constraint.}
    \label{fig:mass_all}
\end{figure*}

\paragraph{Probability Mass.}
We vary the minimum relative probability mass constraint during clustering ($\rho$) from $0.001$ to $0.05$. Enforcing a minimum relative mass ensures clusters receive sufficient supervision, improving learnability for tail items. However, satisfying the constraint may group semantically dissimilar items together, reducing cluster coherence. Figure~\ref{fig:mass_all} shows that no single value is universally optimal: moderate constraints help on some datasets but hurt on others, confirming that this hyperparameter is dataset dependent and should balance semantic quality against clustering balance.

\begin{figure*}[t]
    \centering
    \begin{subfigure}[b]{0.19\textwidth}
        \centering
        \includegraphics[width=\linewidth]{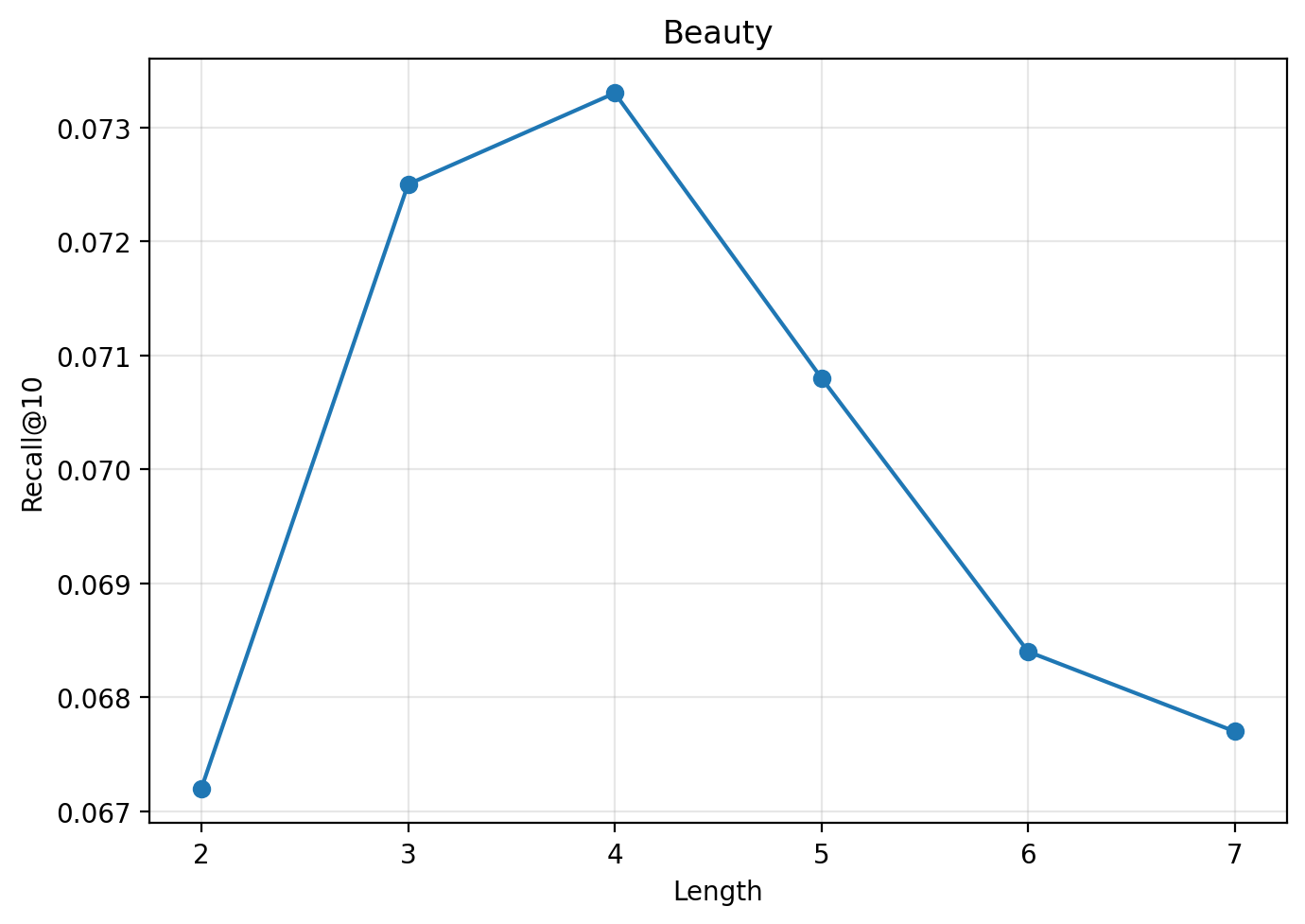}
        \caption{Beauty}
        \label{fig:length_beauty}
    \end{subfigure}
    \hfill
    \begin{subfigure}[b]{0.19\textwidth}
        \centering
        \includegraphics[width=\linewidth]{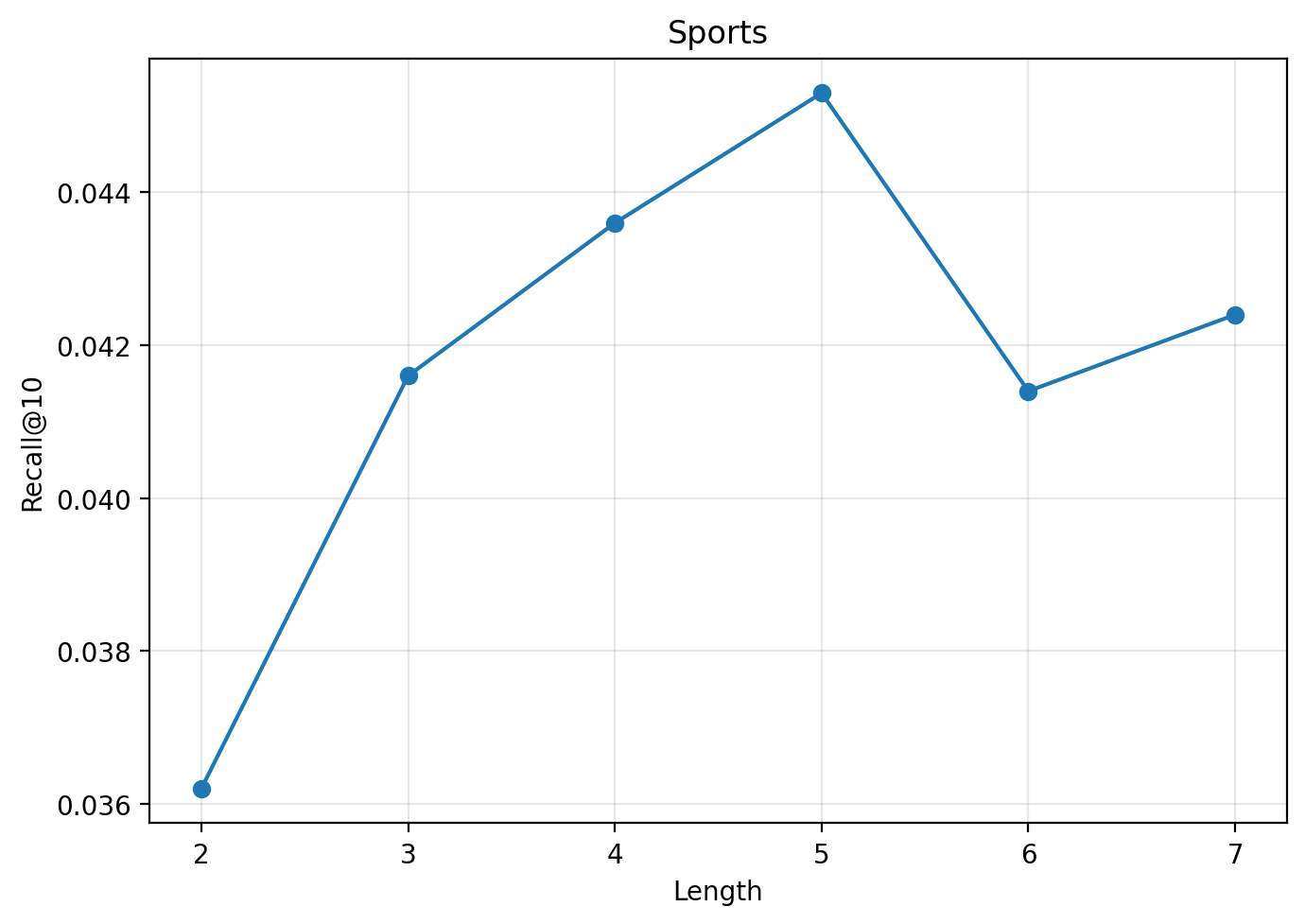}
        \caption{Sports}
        \label{fig:length_sports}
    \end{subfigure}
    \hfill
    \begin{subfigure}[b]{0.19\textwidth}
        \centering
        \includegraphics[width=\linewidth]{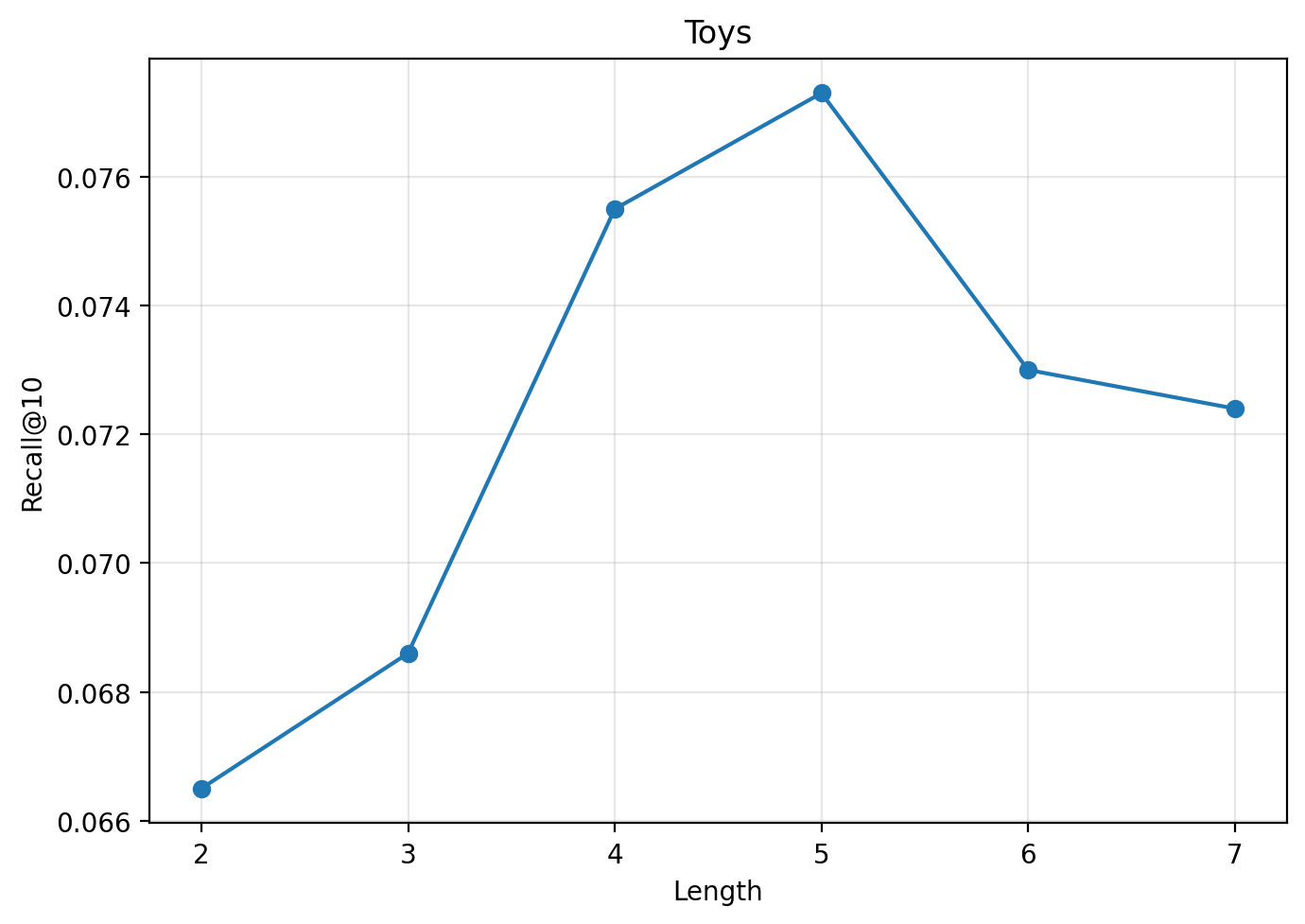}
        \caption{Toys}
        \label{fig:length_toys}
    \end{subfigure}
    \hfill
    \begin{subfigure}[b]{0.19\textwidth}
        \centering
        \includegraphics[width=\linewidth]{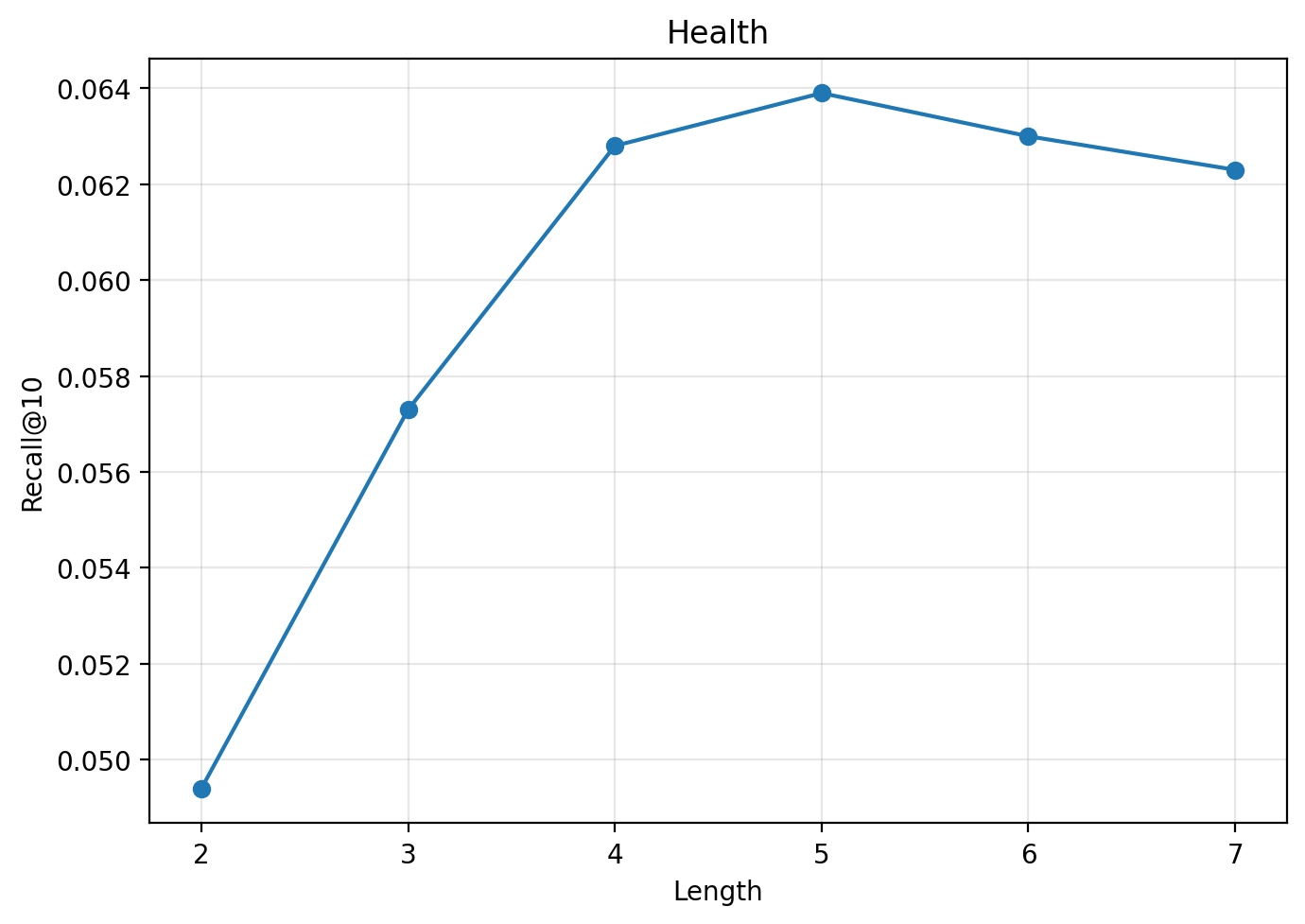}
        \caption{Health}
        \label{fig:length_health}
    \end{subfigure}
    \hfill
    \begin{subfigure}[b]{0.19\textwidth}
        \centering
        \includegraphics[width=\linewidth]{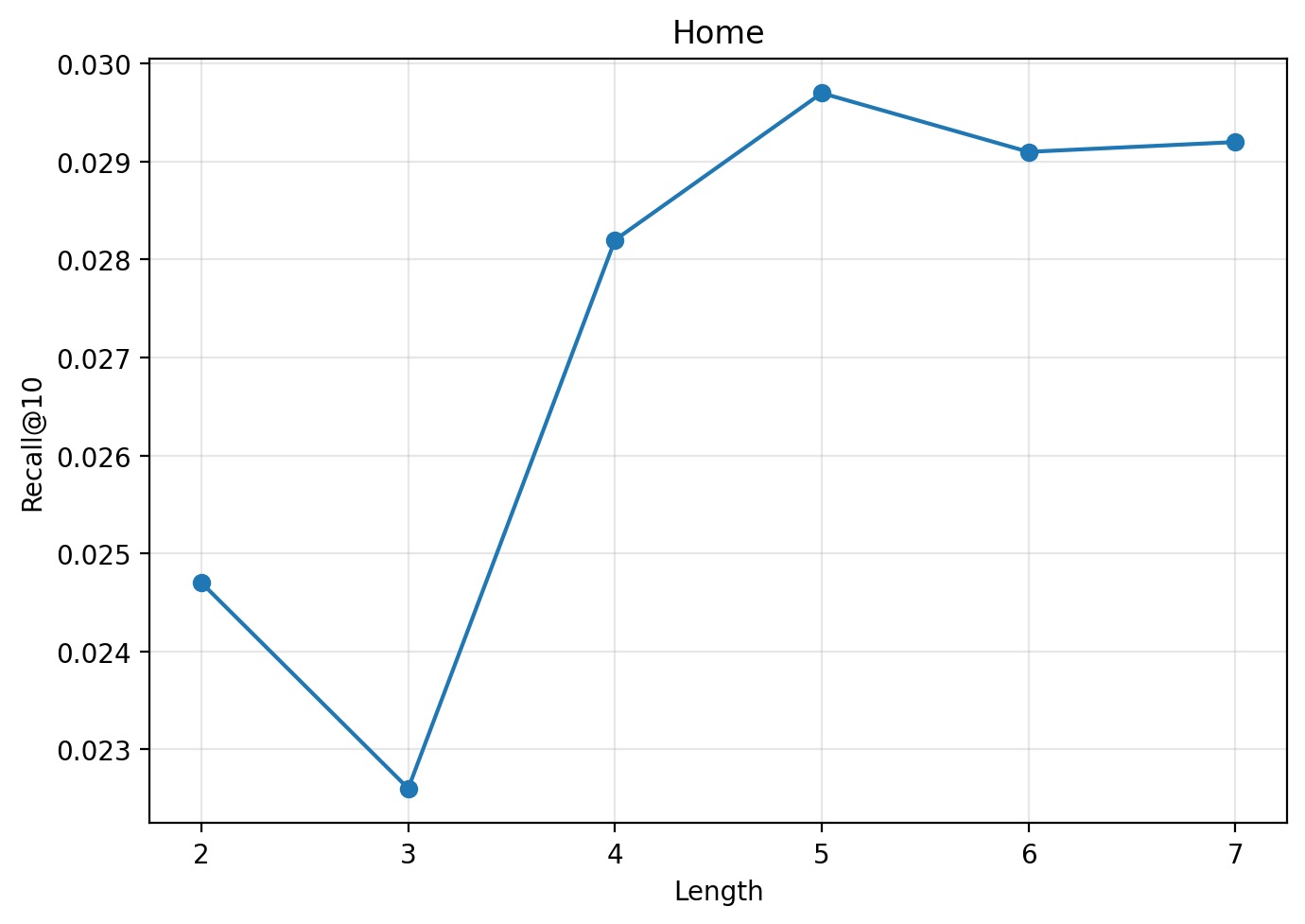}
        \caption{Home}
        \label{fig:length_home}
    \end{subfigure}
    \caption{Recall@10 as a function of semantic ID length.}
    \label{fig:length_all}
\end{figure*}

\paragraph{Length.} We vary the semantic ID length ($m$) from 2 to 7 tokens. Shorter sequences are easier to learn but have limited capacity to encode fine-grained semantics, while longer sequences are more expressive but harder to learn and computationally more expensive. Figure~\ref{fig:length_all} shows that the optimal length is 4 for \textit{Beauty} and 5 for the remaining datasets. Notably, for the larger datasets (\textit{Home} and \textit{Health}), performance remains stable even at lengths 6-7, suggesting that longer sequences require more data to be learned effectively.

\paragraph{Embedding Backbone.}
We keep the downstream pipeline fixed and vary only the embedding model (M3~\cite{chen2025m3}, Qwen3 0.6B/4B) used to construct the semantic IDs. For each dataset, we report AST together with the second-best (SB) and third-best (TB) methods from the main results. Table~\ref{tab:embedding_sensitivity} shows that AST is more stable than competing methods. For example, on Beauty AST varies by only $0.0013$ (relative range $1.75\%$), and on Toys by $0.0013$ (relative range $1.66\%$), while SB and TB exhibit noticeably larger fluctuations. These results suggest that AST is not only strong in absolute performance but also less dependent on the particular embedding backbone.


\begin{table}[t]
\centering
\caption{Sensitivity to the embedding backbone. For each dataset, we report AST with the second-best (SB) and third-best (TB) methods from the main results, varying only the embedding model used to construct semantic IDs.}
\label{tab:embedding_sensitivity}
\setlength{\tabcolsep}{3pt}
\renewcommand{\arraystretch}{0.9}
\small
\begin{tabular}{@{}ll|ccccc@{}}
\toprule
Embedding & Method & Beauty & Toys & Sports & Health & Home \\
\midrule
\multirow{3}{*}{M3}
  & AST & \textbf{7.45} & \textbf{7.84} & \textbf{4.46} & \textbf{6.65} & \textbf{2.80} \\
  & SB  & 7.16 & 7.07 & 4.22 & 5.94 & 2.58 \\
  & TB  & 6.52 & 6.87 & 4.18 & 5.66 & 2.14 \\
\midrule
\multirow{3}{*}{Qwen3-Embedding-0.6B}
  & AST & \textbf{7.51} & \textbf{7.94} & \textbf{4.62} & \textbf{6.64} & \textbf{3.02} \\
  & SB  & 6.98 & 7.07 & 4.09 & 6.30 & 2.63 \\
  & TB  & 6.91 & 6.67 & 4.08 & 6.07 & 2.61 \\
\midrule
\multirow{3}{*}{Qwen3-Embedding-4B}
  & AST & \textbf{7.38} & \textbf{7.81} & \textbf{4.53} & \textbf{6.34} & \textbf{2.91} \\
  & SB  & 6.69 & 6.85 & 4.02 & 5.99 & 2.66 \\
  & TB  & 6.55 & 6.61 & 3.96 & 4.50 & 2.62 \\
\bottomrule
\end{tabular}
\end{table}

\subsection{Ablation Study}
\label{subsec:ablation}

To quantify the contribution of AST's main design choices, we conduct an ablation study along two dimensions. First, we remove the probability-mass constraint by setting $\rho=0$, allowing the hierarchy to split without enforcing a minimum share of probability mass for each child. Second, we disable dynamic branching and replace it with a fixed branching factor applied uniformly across all nodes and levels. For this fixed-branching variant, we perform hyperparameter tuning over the branching factor and report the best-performing configuration for each dataset, ensuring a fair comparison against AST's adaptive mechanism.

The probability-mass constraint avoids extremely small child clusters and maintains sufficiently informative supervision for each semantic prefix. This is crucial in sparse regions of the data, where unconstrained splitting may produce highly imbalanced partitions and weak training signals for low-frequency items. Setting $\rho=0$ gives the clustering procedure full flexibility, but may reduce the robustness of the resulting hierarchy when supervision is limited.
Dynamic branching serves a complementary role. Rather than imposing the same branching factor everywhere, AST adapts the local granularity of the tree to the semantic structure of the data. Nodes with richer internal variation receive finer partitions, while simpler regions remain more compact. Replacing this with a tuned fixed branching factor yields a strong yet rigid hierarchy, allowing us to directly assess the benefit of adaptive branching beyond careful hyperparameter selection.



\begin{table}[t]
\centering
\caption{Ablation study of AST across all datasets. We compare the full model against two variants: (1) removing the probability-mass constraint ($\rho=0$), and (2) replacing dynamic branching with fixed branching. Best results in \textbf{bold}.}
\label{tab:ablation}
\setlength{\tabcolsep}{4pt}
\begin{tabular}{lccc}
\hline
Dataset & AST & \makecell{w/o Mass\\Constraint} & \makecell{w/ Fixed\\Branching} \\
\hline
Beauty  & \textbf{7.51} & 7.33 & 7.01 \\
Toys    & \textbf{7.94} & 7.73 & 7.36 \\
Sports  & \textbf{4.62} & 4.54 & 4.56 \\
Health  & \textbf{6.64} & 6.40 & 6.48 \\
Home    & \textbf{3.02} & 2.98 & 2.95 \\
\hline
\end{tabular}
\end{table}

Table~\ref{tab:ablation} compares the full AST model and ablated variants across datasets. The full model performs best, with balanced mass allocation and adaptive branching improving semantic IDs, while ablated variants remain strong, underscoring AST's effective core bias of modeling semantic IDs as tree paths. A more nuanced pattern emerges across scales. Performance gaps between the full model and its ablations shrink on larger datasets, where abundant interactions provide stronger supervision throughout the hierarchy. On smaller or sparser datasets, the advantages of dynamic branching and mass constraints become more pronounced. This highlights that these components are especially valuable when supervision is scarce: the mass constraint stabilizes prefix-level learning by preventing under-supported branches, while dynamic branching allocates capacity more efficiently across the tree.

\section{Conclusion and Future Work}

We have reimagined semantic IDs as paths in a hierarchical tree rather than independent token sequences. By conditioning each token on its prefix, we reflected a coarse-to-fine partitioning of the item space--a perspective that necessitates a fundamental shift in tokenization design.
We introduced the Adaptive Silhouette Tree (AST), a divisive clustering method that dynamically selects branching factors by maximizing the silhouette coefficient. This allows the tokenizer to adapt to local item densities, producing finer partitions in complex regions. Additionally, our constrained variants incorporate item probability mass to better support long-tail recommendations. Across five Amazon Review datasets, AST consistently outperformed baselines, yielding a significant performance gain.

Despite the promising results, we identify two  limitations that will be addressed in future work. The quality of semantic IDs remains tied to the underlying item representations; noisy or misaligned embeddings can degrade recommendation accuracy. Furthermore, scaling to millions of items requires more efficient clustering and optimization, as the current MILP-based constraints may become computationally expensive. 
Future research should focus on jointly learning representations and tokenizations, developing approximate optimization procedures for better scalability, and evaluating these adaptive structures on larger real-world benchmarks. Addressing these challenges will refine semantic tokenization for industrial-scale generative recommender systems. 


\balance
\bibliographystyle{ACM-Reference-Format}
\bibliography{references}

@article{rajput2023recommender,
  title={Recommender systems with generative retrieval},
  author={Rajput, Shashank and Mehta, Nikhil and Singh, Anima and Hulikal Keshavan, Raghunandan and Vu, Trung and Heldt, Lukasz and Hong, Lichan and Tay, Yi and Tran, Vinh and Samost, Jonah and others},
  journal={Advances in Neural Information Processing Systems},
  volume={36},
  pages={10299--10315},
  year={2023}
}

@misc{denadai2026unifiedlanguagemodellarge,
      title={A Unified Language Model for Large Scale Search, Recommendation, and Reasoning}, 
      author={Marco De Nadai and Edoardo D'Amico and Max Lefarov and Alexandre Tamborrino and Divita Vohra and Mark VanMiddlesworth and Shawn Lin and Jacqueline Wood and Jan Stypka and Eliza Klyce and Keshi Dai and Timothy Christopher Heath and Martin D. Gould and Yves Raimond and Sandeep Ghael and Tony Jebara and Andreas Damianou and Vladan Radosavljevic and Paul N. Bennett and Mounia Lalmas and Praveen Chandar},
      year={2026},
      eprint={2603.17533},
      archivePrefix={arXiv},
      primaryClass={cs.IR},
      url={https://arxiv.org/abs/2603.17533}, 
}

@article{zhao2024recommender,
  title={Recommender systems in the era of large language models (llms)},
  author={Zhao, Zihuai and Fan, Wenqi and Li, Jiatong and Liu, Yunqing and Mei, Xiaowei and Wang, Yiqi and Wen, Zhen and Wang, Fei and Zhao, Xiangyu and Tang, Jiliang and others},
  journal={IEEE Transactions on Knowledge and Data Engineering},
  volume={36},
  number={11},
  pages={6889--6907},
  year={2024},
  publisher={IEEE}
}

@misc{chen2025m3,
      title={M3-Embedding: Multi-Linguality, Multi-Functionality, Multi-Granularity Text Embeddings Through Self-Knowledge Distillation}, 
      author={Jianlv Chen and Shitao Xiao and Peitian Zhang and Kun Luo and Defu Lian and Zheng Liu},
      year={2025},
      eprint={2402.03216},
      archivePrefix={arXiv},
      primaryClass={cs.CL},
      url={https://arxiv.org/abs/2402.03216}, 
}

@misc{penha2025sids,
      title={Semantic IDs for Joint Generative Search and Recommendation}, 
      author={Gustavo Penha and Edoardo D'Amico and Marco De Nadai and Enrico Palumbo and Alexandre Tamborrino and Ali Vardasbi and Max Lefarov and Shawn Lin and Timothy Heath and Francesco Fabbri and Hugues Bouchard},
      year={2025},
      eprint={2508.10478},
      archivePrefix={arXiv},
      primaryClass={cs.IR},
      url={https://arxiv.org/abs/2508.10478}, 
}

@misc{zhang2026gpr,
      title={GPR: Towards a Generative Pre-trained One-Model Paradigm for Large-Scale Advertising Recommendation}, 
      author={Jun Zhang and Yi Li and Yue Liu and Changping Wang and Yuan Wang and Yuling Xiong and Xun Liu and Haiyang Wu and Qian Li and Enming Zhang and Jiawei Sun and Xin Xu and Zishuai Zhang and Ruoran Liu and Suyuan Huang and Zhaoxin Zhang and Zhengkai Guo and Shuojin Yang and Meng-Hao Guo and Huan Yu and Jie Jiang and Shi-Min Hu},
      year={2026},
      eprint={2511.10138},
      archivePrefix={arXiv},
      primaryClass={cs.IR},
      url={https://arxiv.org/abs/2511.10138}, 
}

@inproceedings{wolf2020transformers,
  title={Transformers: State-of-the-art natural language processing},
  author={Wolf, Thomas and Debut, Lysandre and Sanh, Victor and Chaumond, Julien and Delangue, Clement and Moi, Anthony and Cistac, Pierric and Rault, Tim and Louf, R{\'e}mi and Funtowicz, Morgan and others},
  booktitle={Proceedings of the 2020 conference on empirical methods in natural language processing: system demonstrations},
  pages={38--45},
  year={2020}
}

@article{cao2026glass,
  title={GLASS: A Generative Recommender for Long-sequence Modeling via SID-Tier and Semantic Search},
  author={Cao, Shiteng and She, Junda and Liu, Ji and Zeng, Bin and Guo, Chengcheng and Cai, Kuo and Luo, Qiang and Tang, Ruiming and Li, Han and Gai, Kun and others},
  journal={arXiv preprint arXiv:2602.05663},
  year={2026}
}

@article{zhu2025beyond,
  title={Beyond unimodal boundaries: Generative recommendation with multimodal semantics},
  author={Zhu, Jing and Ju, Mingxuan and Liu, Yozen and Koutra, Danai and Shah, Neil and Zhao, Tong},
  journal={arXiv preprint arXiv:2503.23333},
  year={2025}
}

@article{pedregosa2011scikit,
  title={Scikit-learn: Machine learning in Python},
  author={Pedregosa, Fabian and Varoquaux, Ga{\"e}l and Gramfort, Alexandre and Michel, Vincent and Thirion, Bertrand and Grisel, Olivier and Blondel, Mathieu and Prettenhofer, Peter and Weiss, Ron and Dubourg, Vincent and others},
  journal={the Journal of machine Learning research},
  volume={12},
  pages={2825--2830},
  year={2011},
  publisher={JMLR. org}
}

@article{hartigan1979algorithm,
  title={Algorithm AS 136: A k-means clustering algorithm},
  author={Hartigan, John A and Wong, Manchek A},
  journal={Journal of the royal statistical society. series c (applied statistics)},
  volume={28},
  number={1},
  pages={100--108},
  year={1979},
  publisher={JSTOR}
}

@article{raffel2020exploring,
  title={Exploring the limits of transfer learning with a unified text-to-text transformer},
  author={Raffel, Colin and Shazeer, Noam and Roberts, Adam and Lee, Katherine and Narang, Sharan and Matena, Michael and Zhou, Yanqi and Li, Wei and Liu, Peter J},
  journal={Journal of machine learning research},
  volume={21},
  number={140},
  pages={1--67},
  year={2020}
}

@article{zhang2025qwen3,
  title={Qwen3 embedding: Advancing text embedding and reranking through foundation models},
  author={Zhang, Yanzhao and Li, Mingxin and Long, Dingkun and Zhang, Xin and Lin, Huan and Yang, Baosong and Xie, Pengjun and Yang, An and Liu, Dayiheng and Lin, Junyang and others},
  journal={arXiv preprint arXiv:2506.05176},
  year={2025}
}

@inproceedings{he2016ups,
  title={Ups and downs: Modeling the visual evolution of fashion trends with one-class collaborative filtering},
  author={He, Ruining and McAuley, Julian},
  booktitle={proceedings of the 25th international conference on world wide web},
  pages={507--517},
  year={2016}
}

@inproceedings{adamopoulos2014over,
  title={On over-specialization and concentration bias of recommendations: Probabilistic neighborhood selection in collaborative filtering systems},
  author={Adamopoulos, Panagiotis and Tuzhilin, Alexander},
  booktitle={Proceedings of the 8th ACM Conference on Recommender systems},
  pages={153--160},
  year={2014}
}

@article{sreepada2020mitigating,
  title={Mitigating long tail effect in recommendations using few shot learning technique},
  author={Sreepada, Rama Syamala and Patra, Bidyut Kr},
  journal={Expert Systems with Applications},
  volume={140},
  pages={112887},
  year={2020},
  publisher={Elsevier}
}

@article{yin2012challenging,
  title={Challenging the long tail recommendation},
  author={Yin, Hongzhi and Cui, Bin and Li, Jing and Yao, Junjie and Chen, Chen},
  journal={Proceedings of the VLDB Endowment},
  volume={5},
  number={9},
  pages={896--907},
  year={2012},
  publisher={Association for Computing Machinery (ACM)}
}

@article{fang2025hid,
  title={Hid-vae: Interpretable generative recommendation via hierarchical and disentangled semantic ids},
  author={Fang, Dengzhao and Gao, Jingtong and Zhu, Chengcheng and Li, Yu and Zhao, Xiangyu and Chang, Yi},
  journal={arXiv preprint arXiv:2508.04618},
  year={2025}
}

@article{pang2025higr,
  title={HiGR: Efficient Generative Slate Recommendation via Hierarchical Planning and Multi-Objective Preference Alignment},
  author={Pang, Yunsheng and Liu, Zijian and Li, Yudong and Zhu, Shaojie and Luo, Zijian and Yu, Chenyun and Wu, Sikai and Shen, Shichen and Xu, Cong and Wang, Bin and others},
  journal={arXiv preprint arXiv:2512.24787},
  year={2025}
}

@inproceedings{hou2025generating,
  title={Generating long semantic ids in parallel for recommendation},
  author={Hou, Yupeng and Li, Jiacheng and Shin, Ashley and Jeon, Jinsung and Santhanam, Abhishek and Shao, Wei and Hassani, Kaveh and Yao, Ning and McAuley, Julian},
  booktitle={Proceedings of the 31st ACM SIGKDD Conference on Knowledge Discovery and Data Mining V. 2},
  pages={956--966},
  year={2025}
}

@article{tang2025reaseq,
  title={ReaSeq: Unleashing World Knowledge via Reasoning for Sequential Modeling},
  author={Tang, Jiakai and Wang, Chuan and Yang, Gaoming and Wu, Han and Yu, Jiahao and Wu, Jian and Hu, Jianwu and Zheng, Junjun and Li, Longbin and Xiao, Shuwen and others},
  journal={arXiv preprint arXiv:2512.21257},
  year={2025}
}

@article{he2025plum,
  title={Plum: Adapting pre-trained language models for industrial-scale generative recommendations},
  author={He, Ruining and Heldt, Lukasz and Hong, Lichan and Keshavan, Raghunandan and Mao, Shifan and Mehta, Nikhil and Su, Zhengyang and Tsai, Alicia and Wang, Yueqi and Wang, Shao-Chuan and others},
  journal={arXiv preprint arXiv:2510.07784},
  year={2025}
}

@article{kong2025minionerec,
  title={Minionerec: An open-source framework for scaling generative recommendation},
  author={Kong, Xiaoyu and Sheng, Leheng and Tan, Junfei and Chen, Yuxin and Wu, Jiancan and Zhang, An and Wang, Xiang and He, Xiangnan},
  journal={arXiv preprint arXiv:2510.24431},
  year={2025}
}

@inproceedings{lin2025order,
  title={Order-agnostic identifier for large language model-based generative recommendation},
  author={Lin, Xinyu and Shi, Haihan and Wang, Wenjie and Feng, Fuli and Wang, Qifan and Ng, See-Kiong and Chua, Tat-Seng},
  booktitle={Proceedings of the 48th international ACM SIGIR conference on research and development in information retrieval},
  pages={1923--1933},
  year={2025}
}

@inproceedings{hou2025actionpiece,
  title={ActionPiece: Contextually Tokenizing Action Sequences for Generative Recommendation},
  author={Hou, Yupeng and Ni, Jianmo and He, Zhankui and Sachdeva, Noveen and Kang, Wang-Cheng and Chi, Ed H and Mcauley, Julian and Cheng, Derek Zhiyuan},
  booktitle={International Conference on Machine Learning},
  pages={24004--24024},
  year={2025},
  organization={PMLR}
}

@article{petrov2023generative,
  title={Generative sequential recommendation with gptrec},
  author={Petrov, Aleksandr V and Macdonald, Craig},
  journal={arXiv preprint arXiv:2306.11114},
  year={2023}
}

@inproceedings{tan2024idgenrec,
  title={Idgenrec: Llm-recsys alignment with textual id learning},
  author={Tan, Juntao and Xu, Shuyuan and Hua, Wenyue and Ge, Yingqiang and Li, Zelong and Zhang, Yongfeng},
  booktitle={Proceedings of the 47th international ACM SIGIR conference on research and development in information retrieval},
  pages={355--364},
  year={2024}
}

@article{rousseeuw1987silhouettes,
  title={Silhouettes: a graphical aid to the interpretation and validation of cluster analysis},
  author={Rousseeuw, Peter J},
  journal={Journal of computational and applied mathematics},
  volume={20},
  pages={53--65},
  year={1987},
  publisher={Elsevier}
}

@inproceedings{wang2024content,
  title={Content-based collaborative generation for recommender systems},
  author={Wang, Yidan and Ren, Zhaochun and Sun, Weiwei and Yang, Jiyuan and Liang, Zhixiang and Chen, Xin and Xie, Ruobing and Yan, Su and Zhang, Xu and Ren, Pengjie and others},
  booktitle={Proceedings of the 33rd ACM International Conference on Information and Knowledge Management},
  pages={2420--2430},
  year={2024}
}

@inproceedings{hua2023index,
  title={How to index item ids for recommendation foundation models},
  author={Hua, Wenyue and Xu, Shuyuan and Ge, Yingqiang and Zhang, Yongfeng},
  booktitle={Proceedings of the Annual International ACM SIGIR Conference on Research and Development in Information Retrieval in the Asia Pacific Region},
  pages={195--204},
  year={2023}
}

@inproceedings{si2024generative,
  title={Generative retrieval with semantic tree-structured identifiers and contrastive learning},
  author={Si, Zihua and Sun, Zhongxiang and Chen, Jiale and Chen, Guozhang and Zang, Xiaoxue and Zheng, Kai and Song, Yang and Zhang, Xiao and Xu, Jun and Gai, Kun},
  booktitle={Proceedings of the 2024 Annual International ACM SIGIR Conference on Research and Development in Information Retrieval in the Asia Pacific Region},
  pages={154--163},
  year={2024}
}

@article{vallaeys2025qinco2,
  title={Qinco2: Vector compression and search with improved implicit neural codebooks},
  author={Vallaeys, Th{\'e}ophane and Muckley, Matthew and Verbeek, Jakob and Douze, Matthijs},
  journal={arXiv preprint arXiv:2501.03078},
  year={2025}
}

@article{zhang2025purely,
  title={Purely Semantic Indexing for LLM-based Generative Recommendation and Retrieval},
  author={Zhang, Ruohan and Li, Jiacheng and McAuley, Julian and Hou, Yupeng},
  journal={arXiv preprint arXiv:2509.16446},
  year={2025}
}

@article{wang2022neural,
  title={A neural corpus indexer for document retrieval},
  author={Wang, Yujing and Hou, Yingyan and Wang, Haonan and Miao, Ziming and Wu, Shibin and Chen, Qi and Xia, Yuqing and Chi, Chengmin and Zhao, Guoshuai and Liu, Zheng and others},
  journal={Advances in Neural Information Processing Systems},
  volume={35},
  pages={25600--25614},
  year={2022}
}

@inproceedings{de2020autoregressive,
  title={Autoregressive Entity Retrieval},
  author={De Cao, N and Izacard, G and Riedel, S and Petroni, F},
  booktitle={ICLR 2021-9th International Conference on Learning Representations},
  volume={2021},
  year={2020},
  organization={ICLR}
}

@article{defossez2022high,
  title={High fidelity neural audio compression},
  author={D{\'e}fossez, Alexandre and Copet, Jade and Synnaeve, Gabriel and Adi, Yossi},
  journal={arXiv preprint arXiv:2210.13438},
  year={2022}
}

@article{yu2023language,
  title={Language Model Beats Diffusion--Tokenizer is Key to Visual Generation},
  author={Yu, Lijun and Lezama, Jos{\'e} and Gundavarapu, Nitesh B and Versari, Luca and Sohn, Kihyuk and Minnen, David and Cheng, Yong and Birodkar, Vighnesh and Gupta, Agrim and Gu, Xiuye and others},
  journal={arXiv preprint arXiv:2310.05737},
  year={2023}
}

@inproceedings{esser2021taming,
  title={Taming transformers for high-resolution image synthesis},
  author={Esser, Patrick and Rombach, Robin and Ommer, Bjorn},
  booktitle={Proceedings of the IEEE/CVF conference on computer vision and pattern recognition},
  pages={12873--12883},
  year={2021}
}

@article{razavi2019generating,
  title={Generating diverse high-fidelity images with vq-vae-2},
  author={Razavi, Ali and Van den Oord, Aaron and Vinyals, Oriol},
  journal={Advances in neural information processing systems},
  volume={32},
  year={2019}
}

@article{van2017neural,
  title={Neural discrete representation learning},
  author={Van Den Oord, Aaron and Vinyals, Oriol and others},
  journal={Advances in neural information processing systems},
  volume={30},
  year={2017}
}

@inproceedings{zeng2024scalable,
  title={Scalable and effective generative information retrieval},
  author={Zeng, Hansi and Luo, Chen and Jin, Bowen and Sarwar, Sheikh Muhammad and Wei, Tianxin and Zamani, Hamed},
  booktitle={Proceedings of the ACM Web Conference 2024},
  pages={1441--1452},
  year={2024}
}

@article{tay2022transformer,
  title={Transformer memory as a differentiable search index},
  author={Tay, Yi and Tran, Vinh and Dehghani, Mostafa and Ni, Jianmo and Bahri, Dara and Mehta, Harsh and Qin, Zhen and Hui, Kai and Zhao, Zhe and Gupta, Jai and others},
  journal={Advances in neural information processing systems},
  volume={35},
  pages={21831--21843},
  year={2022}
}

@article{rubenstein2023audiopalm,
  title={Audiopalm: A large language model that can speak and listen},
  author={Rubenstein, Paul K and Asawaroengchai, Chulayuth and Nguyen, Duc Dung and Bapna, Ankur and Borsos, Zal{\'a}n and Quitry, F{\'e}lix de Chaumont and Chen, Peter and Badawy, Dalia El and Han, Wei and Kharitonov, Eugene and others},
  journal={arXiv preprint arXiv:2306.12925},
  year={2023}
}

@article{borsos2023audiolm,
  title={Audiolm: a language modeling approach to audio generation},
  author={Borsos, Zal{\'a}n and Marinier, Rapha{\"e}l and Vincent, Damien and Kharitonov, Eugene and Pietquin, Olivier and Sharifi, Matt and Roblek, Dominik and Teboul, Olivier and Grangier, David and Tagliasacchi, Marco and others},
  journal={IEEE/ACM transactions on audio, speech, and language processing},
  volume={31},
  pages={2523--2533},
  year={2023},
  publisher={IEEE}
}

@article{hadad2026encoderec,
  title={EncodeRec: An Embedding Backbone for Recommendation Systems},
  author={Hadad, Guy and Rabaev, Neomi and Shapira, Bracha},
  journal={arXiv preprint arXiv:2601.10837},
  year={2026}
}

@article{tian2024visual,
  title={Visual autoregressive modeling: Scalable image generation via next-scale prediction},
  author={Tian, Keyu and Jiang, Yi and Yuan, Zehuan and Peng, Bingyue and Wang, Liwei},
  journal={Advances in neural information processing systems},
  volume={37},
  pages={84839--84865},
  year={2024}
}

@inproceedings{lee2022autoregressive,
  title={Autoregressive image generation using residual quantization},
  author={Lee, Doyup and Kim, Chiheon and Kim, Saehoon and Cho, Minsu and Han, Wook-Shin},
  booktitle={Proceedings of the IEEE/CVF conference on computer vision and pattern recognition},
  pages={11523--11532},
  year={2022}
}

@article{lloyd1982least,
  title={Least squares quantization in PCM},
  author={Lloyd, Stuart},
  journal={IEEE Transactions on Information Theory},
  volume={28},
  number={2},
  pages={129--137},
  year={1982},
  publisher={IEEE}
}

@article{wu2024survey,
  title={A survey on large language models for recommendation},
  author={Wu, Likang and Zheng, Zhi and Qiu, Zhaopeng and Wang, Hao and Gu, Hongchao and Shen, Tingjia and Qin, Chuan and Zhu, Chen and Zhu, Hengshu and Liu, Qi and others},
  journal={World Wide Web},
  volume={27},
  number={5},
  pages={60},
  year={2024},
  publisher={Springer}
}

@article{vaswani2017attention,
  title={Attention is all you need},
  author={Vaswani, Ashish and Shazeer, Noam and Parmar, Niki and Uszkoreit, Jakob and Jones, Llion and Gomez, Aidan N and Kaiser, {\L}ukasz and Polosukhin, Illia},
  journal={Advances in neural information processing systems},
  volume={30},
  year={2017}
}

@inproceedings{zheng2024adapting,
  title={Adapting large language models by integrating collaborative semantics for recommendation},
  author={Zheng, Bowen and Hou, Yupeng and Lu, Hongyu and Chen, Yu and Zhao, Wayne Xin and Chen, Ming and Wen, Ji-Rong},
  booktitle={2024 IEEE 40th International Conference on Data Engineering (ICDE)},
  pages={1435--1448},
  year={2024},
  organization={IEEE}
}

@article{yang2025qwen3,
  title={Qwen3 technical report},
  author={Yang, An and Li, Anfeng and Yang, Baosong and Zhang, Beichen and Hui, Binyuan and Zheng, Bo and Yu, Bowen and Gao, Chang and Huang, Chengen and Lv, Chenxu and others},
  journal={arXiv preprint arXiv:2505.09388},
  year={2025}
}

@inproceedings{wang2024learnable,
  title={Learnable item tokenization for generative recommendation},
  author={Wang, Wenjie and Bao, Honghui and Lin, Xinyu and Zhang, Jizhi and Li, Yongqi and Feng, Fuli and Ng, See-Kiong and Chua, Tat-Seng},
  booktitle={Proceedings of the 33rd ACM International Conference on Information and Knowledge Management},
  pages={2400--2409},
  year={2024}
}

@article{deng2025onerec,
  title={Onerec: Unifying retrieve and rank with generative recommender and iterative preference alignment},
  author={Deng, Jiaxin and Wang, Shiyao and Cai, Kuo and Ren, Lejian and Hu, Qigen and Ding, Weifeng and Luo, Qiang and Zhou, Guorui},
  journal={arXiv preprint arXiv:2502.18965},
  year={2025}
}

@inproceedings{ju2025generative,
  title={Generative Recommendation with Semantic IDs: A Practitioner's Handbook},
  author={Ju, Clark Mingxuan and Collins, Liam and Neves, Leonardo and Kumar, Bhuvesh and Wang, Louis Yufeng and Zhao, Tong and Shah, Neil},
  booktitle={Proceedings of the 34th ACM International Conference on Information and Knowledge Management},
  pages={6420--6425},
  year={2025}
}

@inproceedings{wang2024eager,
  title={Eager: Two-stream generative recommender with behavior-semantic collaboration},
  author={Wang, Ye and Xun, Jiahao and Hong, Minjie and Zhu, Jieming and Jin, Tao and Lin, Wang and Li, Haoyuan and Li, Linjun and Xia, Yan and Zhao, Zhou and others},
  booktitle={Proceedings of the 30th ACM SIGKDD Conference on Knowledge Discovery and Data Mining},
  pages={3245--3254},
  year={2024}
}

@inproceedings{zhu2024cost,
  title={Cost: Contrastive quantization based semantic tokenization for generative recommendation},
  author={Zhu, Jieming and Jin, Mengqun and Liu, Qijiong and Qiu, Zexuan and Dong, Zhenhua and Li, Xiu},
  booktitle={Proceedings of the 18th ACM Conference on Recommender Systems},
  pages={969--974},
  year={2024}
}

\end{document}